\documentclass[a4paper, oneside, 12pt]{article}      
\usepackage[Lenny]{fncychap}

\usepackage[utf8]{inputenc}
\usepackage[english]{babel}
\usepackage{verbatim}
\usepackage[show]{chato-notes}
\usepackage{enumitem}
\usepackage{booktabs}
\usepackage{adjustbox}
\usepackage{cprotect}
\usepackage{hyperref}
\usepackage{amsmath}
\usepackage{longtable}

\usepackage{mathpazo}
\usepackage{inconsolata}

\newcommand{\ifigure}[2]{
	{\par\centering
	\resizebox*{#1\textwidth}{!}{\includegraphics{#2}}
	\par}
}

\newcommand{\wordtovec}{\ensuremath{\sf word2vec}}
\newcommand{\fasttext}{\ensuremath{\sf fasttext}}
\newcommand{\glove}{\ensuremath{\sf GloVe}}

\newcommand{\drmm}{\ensuremath{\sf DRMM}}
\newcommand{\knrm}{\ensuremath{\sf KNRM}}
\newcommand{\convknrm}{\ensuremath{\sf ConvKNRM}}
\newcommand{\pacrr}{\ensuremath{\sf PACRR}}

\newcommand{\cls}{\texttt{[CLS]}}
\newcommand{\sep}{\texttt{[SEP]}}

\newcommand{\out}{\texttt{[OUT]}}
\newcommand{\softmax}{\ensuremath{\sf softmax}}

\newcommand{\encoder}{\ensuremath{\sf Encoder}}
\newcommand{\relu}{\ensuremath{\sf ReLU}}

\newcommand{\bert}{\ensuremath{\sf BERT}}
\newcommand{\roberta}{\ensuremath{\sf RoBERTa}}
\newcommand{\distilbert}{\ensuremath{\sf DistilBERT}}
\newcommand{\bart}{\ensuremath{\sf BART}}
\newcommand{\tfive}{\ensuremath{\sf T5}}
\newcommand{\gpt}{\ensuremath{\sf GPT}}

\newcommand{\firstp}{\ensuremath{\sf FirstP}}
\newcommand{\maxp}{\ensuremath{\sf MaxP}}
\newcommand{\sump}{\ensuremath{\sf SumP}}

\newcommand{\dpr}{\ensuremath{\sf DPR}}
\newcommand{\ance}{\ensuremath{\sf ANCE}}
\renewcommand{\star}{\ensuremath{\sf STAR}}

\newcommand{\colbert}{\ensuremath{\sf ColBERT}}
\newcommand{\mebert}{\ensuremath{\sf ME\text{-}\bert}}
\newcommand{\coil}{\ensuremath{\sf COIL}}
\newcommand{\blas}{\ensuremath{\sf BLAS}}

\newcommand{\dtfiveq}{\ensuremath{\sf DocT5Query}}
\newcommand{\dtoq}{\ensuremath{\sf Doc2Query}}
\renewcommand{\tilde}{\ensuremath{\sf TILDEv2}}
\newcommand{\sparterm}{\ensuremath{\sf SparTerm}}
\newcommand{\deepct}{\ensuremath{\sf DeepCT}}
\newcommand{\deepimpact}{\ensuremath{\sf DeepImpact}}
\newcommand{\unicoil}{\ensuremath{\sf UniCOIL}}
\newcommand{\splade}{\ensuremath{\sf SPLADE}}
\newcommand{\flops}{\ensuremath{\sf FLOPS}}
\newcommand{\epic}{\ensuremath{\sf EPIC}}

\usepackage{fullpage}
\usepackage{setspace}
\usepackage[authoryear,sort,square]{natbib}
\bibpunct{[}{]}{,}{a}{}{,}    % Change fences to brackets;
\let \cite = \citep

\newcommand{\chapter}[1]{}
\newcommand{\chapterauthor}[1]{}

\newcommand{\HRule}[1]{%
  \nointerlineskip \vspace{\baselineskip}%
  \hspace{\fill}\rule{0.5\linewidth}{#1}\hspace{\fill}%
  \par\nointerlineskip \vspace{\baselineskip}
}

\definecolor{grey}{rgb}{0.9,0.9,0.9} 

\makeatletter							% Title
\def\printtitle{%						
    {\centering \@title}}
\makeatother									

\makeatletter							% Author
\def\printauthor{%					
    {\centering \large \@author}}				
\makeatother							

\makeatletter							% Date
\def\printdate{%					
    {\centering \large \@date}}				
\makeatother

\begin{document}

% ------------------------------------------------------------------------------
% Metadata (Change this)
% ------------------------------------------------------------------------------
\title{	\fontsize{30}{60}\selectfont
			\vspace*{0.7cm}
			\hfill Lecture Notes on \\[0.8cm]
			%\hfill per 	\\[0.8cm]
			\hfill Neural Information Retrieval%
		}

\author{
		\hfill Nicola Tonellotto\\
		\hfill \url{nicola.tonellotto@unipi.it}\\
}
    
\date{
	\hfill Version date: \today\\
	\hfill Copyright \copyright 2022. All rights reserved.\\
}
% ------------------------------------------------------------------------------
% Maketitle
% ------------------------------------------------------------------------------
\thispagestyle{empty}				% Remove page numbering on this page

\colorbox{grey}{
	\parbox[t]{1.0\linewidth}{
		\printtitle 
		\vspace*{0.7cm}
	}
}

\vfill
\printauthor								% Print the author data as defined above
\HRule{1pt}

\printdate

\clearpage
\newpage

These lecture notes focus on the recent advancements in neural information retrieval, with particular emphasis on the systems and models exploiting transformer networks. These networks, originally proposed by Google in 2017, have seen a large success in many natural language processing and information retrieval tasks. While there are many fantastic textbook on information retrieval~\cite{manning:2008,buttcher:2010} and natural language processing~\cite{Jurafsky2009}, as well as specialised books for a more advanced audience~\cite{fntir,Liu09ftir,barla:2015,lin2020pretrained}, these lecture notes target people aiming at developing a basic understanding of the main information retrieval techniques and approaches based on deep learning.

\medskip

These notes have been prepared for a graduate course of the MSc program in Artificial Intelligence and Data Engineering at the University of Pisa, Italy. Part of the material is inspired by other works, and when this is done a reference is obviously provided to the original. I would like to warmly thank all the students and colleagues who read these notes, gave me their feedback and sent me their corrections, that allowed to fix many errors on the original manuscript. 

\medskip

I am aware that this document is far from perfect, and I am eager to improve it. Feel free to contact me if you have any comments, suggestions and/or if you find typos/mistakes.

\subsubsection*{Version History}

\begin{description}[itemsep=-3mm]
    \item[27/07/2022] Initial version.
    \item[12/09/2022] Fixed typos and terminology.
\end{description}

\newpage

\tableofcontents

\newpage

\vspace*{\fill}
\hfill
\begin{center}
This page intentionally left blank.
\end{center}
\vspace{\fill}

\section*{Introduction}
\chapter{Neural Information Retrieval}
\chapterauthor{Nicola Tonellotto\\University of Pisa}

%\section*{Introduction}
Text Information Retrieval (IR) systems focus on retrieving text documents able to fulfill the information needs of their users, typically expressed as textual queries. Over the years, this inherently vague description has been formalised and characterised by the specific nature of documents, information needs, and users. At the core of the formalisation lies the concept of relevance of a document with respect to a query, and how to estimate their relevance. Over the years, many different ranking models have been proposed to estimate the relevance of documents in response to a query. These models depend on the information provided by the queries and the documents, that are exploited to derive ``relevance signals''. Many ranking models have been developed over the years, ranging from Boolean models to probabilistic and statistical language models. These ``bag of words'' models leverage the presence or the number of occurrences of query terms in the documents to infer their relevance to a query, exploiting hand-crafted functions to combine these occurrences such as BM25.  With the rise of the Web and social platforms, more sources of relevance information about documents have been identified. Machine learning methods have been proved effective to deal with this abundance of relevance signals, and their application to rank the documents in order of relevance estimates w.r.t. a query has given birth to many learning-to-rank (LTR) models. Relevance signals are input features in LTR models, and they are often designed by hand, a time-consuming process. Motivated by their breakthroughs in many computer vision and natural language processing tasks, neural networks represent the current state-of-the-art approach in ranking documents w.r.t. query relevance. 

Neural Information Retrieval focuses on retrieving text documents able to fulfill the information needs of their users exploiting deep neural networks. In neural IR, neural networks are typically used in two different ways: to learn the ranking functions combining the relevance signals to produce an ordering of documents, and to learn the abstract representations of documents and queries to capture their relevance information. In the following, we provide an introduction to the recent approaches in neural IR. Since the research in the field is rapidly evolving, we do not pretend to cover every single aspect of neural IR, but to provide a principled introduction to the main ideas and existing systems in the field. When available, we provide links to relevant and more detailed surveys.

%\mynote{quello che segue deve essere completamente cambiato, man mano che mi viene in mente la struttura migliore.}
Here is a quick overview over what the sections are about.
Section~\ref{sec:math} provides a short depiction of the different representations for text adopted in IR, from the classical BOW encodings to learning-to-rank features to word embeddings.
Section~\ref{sec:interaction} presents the main neural architectures for computing a joint representation of query and document pairs for relevance ranking.
Section~\ref{sec:representation} focuses on the neural architectures specifically tailored for learning abstract complex representations of query and documents texts independently.
Section~\ref{sec:dense} overviews the deployment schemes adopted in neural IR systems, together with an overview of the most common dense retrieval indexes supporting exact and approximate nearest neighbour search.
Section~\ref{sec:sparse} discusses the current approaches in learned sparse retrieval, dealing with the learning of low dimensional representations of documents amenable to be stored in an inverted index or similar data structures.
Finally, Section~\ref{sec:conc} draws concluding remarks.

\subsection*{Notes}

In the following, we will refer to the textual contents of queries and documents as \textit{text}, and we will refer to long documents and short documents, also known as passages, as \textit{documents},  according to common practice in  neural IR. In IR, tokenisation algorithms, implemented by tokenisers, decompose both query and document texts into atomic units of text called \textit{tokens}. The atomic units of text can represent whole \textit{words}, e.g., \texttt{username} or \textit{sub-words}, e.g., \texttt{user\#\#} and \texttt{\#\#name}, depending on the tokenisation algorithm employed to decompose them. The collection of unique tokens in a text collection is called a \textit{vocabulary}, whose elements are called \textit{terms}. These terms in turn correspond to words or sub-words in queries and documents, depending on the tokenisation algorithm.

Moreover, for simplicity, we will describe the linear neural layer $y = Ax + b$, where $A$ is the weight matrix and $b$ is the bias vector, using the simplified notation $y = Wx$, representing exactly the same layer without explicitly referring to the bias vector.

\newpage
\section*{List of Symbols}

\begin{longtable}{p{0.15\textwidth}p{0.75\textwidth}}
$\mathbb{N}$ & the set of natural numbers \\
$\mathbb{R}$ & the set of real numbers \\
$\mathbb{R}^n$ & the set of vectors, or \textit{embeddings}, that consist of $n$ real-valued entries\\
$V_1, V_2, V_3$ & latent representation spaces \\
$V$ & a vector space \\
$w$ & a term (word or sub-word) \\
$t$ & a text, composed by tokens (words or sub-words) \\
$d$ & a text document, composed by tokens (words or sub-words)\\
$q$ & a text query, composed by tokens (words or sub-words)\\
$\mathcal{V}$ & a vocabulary, whose elements are terms \\
$\mathcal{D}$ & a document collection, whose elements are documents \\
$\mathcal{Q}$ & a query log, whose elements are queries\\
$\phi(q), \phi$ & an embedding of the query $q$\\
$\psi(d), \psi$ & an embedding of the document $d$\\
$\eta(q, d)$ & an embedding of the query-document pair $q,d$\\
$\phi_1, \phi_2, \ldots$ & components of a query embedding or generic query embeddings\\
$\psi_1, \psi_2, \ldots$ & component of a document embedding or generic document embeddings\\
$s(q,d)$ & the relevance score of the document $d$ w.r.t. the query $q$\\
$f(\cdot)$ & an embedding aggregation function\\
$k,m,n$ & positive integer values\\
$\ell$ & dimension of output embeddings of transformer architectures, e.g., \bert\ and \tfive\ models; if not otherwise specified, $\ell = 768$\\
\cls & a special classification token in \bert\ and \tfive\ models\\
\sep & a special separation token in \bert\ and \tfive\ models\\
\out & a special output token in \tfive\ models\\
$q_1, \ldots, q_m$ & query tokens\\
$d_1, \ldots, d_n$ & document tokens\\
%$f()$ & a generic aggregation function\\

\end{longtable}
\newpage

\section{Text Representations for Ranking}\label{sec:math}

According to the Probability Ranking Principle~\cite{prp}, under certain assumptions, for a given user's query, documents in a collection should be ranked in order of the (decreasing) probability of relevance w.r.t. the query, to maximise the overall effectiveness of a retrieval system for the user. The task of \textit{ad-hoc ranking} is, for each query, to compute an ordering of the documents equivalent or most similar to the optimal ordering based on the probability of relevance. It is common to limit the documents to be ordered to just the top $k$ documents in the optimal ordering.

Let $\mathcal{D}$ denote a collection of (text) documents, and $\mathcal{Q}$ denote a log of (text) queries. Queries and documents share the same vocabulary $\mathcal{V}$ of terms. A \textit{ranking function}, also known as \textit{scoring function}, $s: \mathcal{Q} \times \mathcal{D} \to \mathbb{R}$ computes a real-valued score for the documents in the collection $\mathcal{D}$ w.r.t. the queries in the log $\mathcal{Q}$. Given a query $q$ and a document $d$, we call the value $s(q,d)$ \textit{relevance score} of the document w.r.t. the query. For a given query, the scores of the documents in the collection can be used to induce an ordering of the documents, in reverse value of score. The closer this induced ordering is to the optimal ordering, the more effective an IR system based on the scoring function is.

Without loss of generality, the scoring function $s(q,d)$ can be further decomposed as:
\begin{equation}\label{eq:scoring}
    s(q,d) = f\big(\phi(q), \psi(d), \eta(q,d)\big),
\end{equation}
where $\phi: \mathcal{Q} \to V_1$, $\psi: \mathcal{D} \to V_2$, and $\eta: \mathcal{Q} \times \mathcal{D} \to V_3$ are three \textit{representation functions}, mapping queries, documents, and query-document pairs into the \textit{latent representation spaces} $V_1$, $V_2$, and $V_3$, respectively~\cite{GUO2020102067}. These functions build abstract mathematical representations of the text sequences of documents and queries amenable for computations. 
%In the vector-space model (VSM)~\cite{vsm}, the representation spaces are assumed to be identical finite-dimensional vector spaces, coincident with $\mathbb{R}^d$. 
\looseness -1 The elements of these vectors represent the features used to describe the corresponding objects, and the \textit{aggregation function} $f: V_1 \times V_2 \times V_3 \to \mathbb{R}$ computes the relevance score of the document representation w.r.t. the query representation.

\begin{figure}[htb!]
    \ifigure{.90}{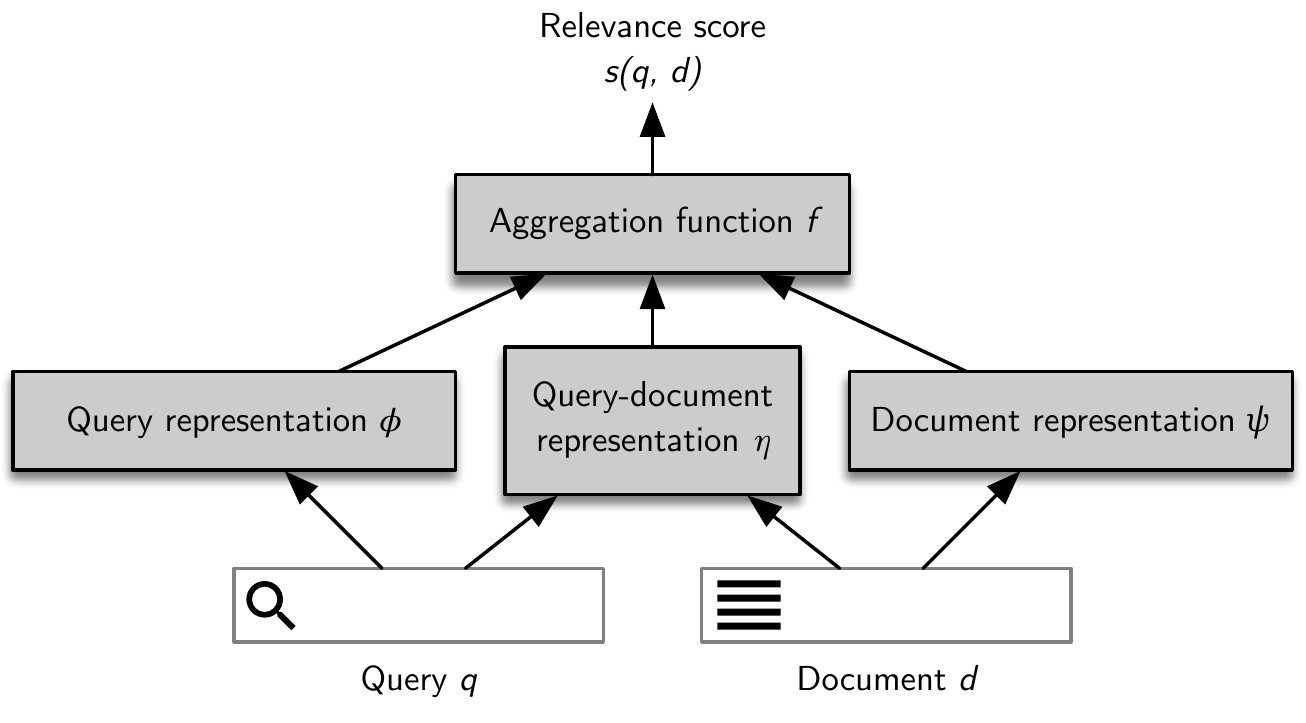}
    \caption{Representation-based decomposition of a ranking function. Adapted from~\cite{GUO2020102067}.}\label{fig:representations}
\end{figure}

The representation functions $\phi$, $\psi$ and $\eta$, and the aggregation function $f$ can be designed by hand, leveraging some axioms or heuristics, or computed through machine learning algorithms. In the following, we will overview the representation functions adopted in classical IR (Section~\ref{ssec:onehot}), in LTR scenarios (Section~\ref{ssec:ltr}) and the recently proposed word embeddings (Section~\ref{ssec:wordemb}).

\subsection{BOW Encodings}\label{ssec:onehot}

In classical IR, both representation and aggregation functions are designed manually, incorporating some lexical statistics such as number of occurrences of terms in a document or in the whole collection. 
Classical IR ranking models, e.g., vector space models~\cite{vsm}, probabilistic models~\cite{bm25} and statistical language models~\cite{lm}, are based on the \textit{bag of words} (BOW) model, where queries and documents are represented as a set of terms from the vocabulary $\mathcal{V}$ together with the number of occurrences of the corresponding tokens in the text. 
More formally, queries and documents are represented as vectors $\phi(q)$ and $\psi(d)$ in $\mathbb{N}^{|\mathcal{V}|}$, called \textit{BOW encodings}, where the $i$-th component of both representations encodes the number of occurrences of the term $w_i \in \mathcal{V}$ in the corresponding text. The query-document representation function $\eta$ is not present in these ranking functions.
%These ranking functions have the form
%\begin{equation}\label{eq:ranking:classical}
%s(q,d) = \sum_{t \in q} tf_q(t) s(t,d),
%\end{equation}
%where the sum is computed over the terms appearing at least once in the query, the quantity $tf_q(t)$, called in-query term %frequency, is the number of occurrences of term $t$ in the query, and the quantity $s(t,d)$ represents the score of the %singe term $t$ appearing in the document $d$.
%
% In these models, queries and documents are represented as vectors $\phi(q)$ and $\psi(d)$ in $\mathbb{R}^{\mathcal{V}}$, where the $i$-th component of both representations encodes information about the term $t_i \in \mathcal{V}$. Hence, the value $\phi_i(q)$ represents the in-query term frequency of the term $t_i$, and the value $\psi_i(d)$ represents the term-document score $s(t_i,d)$; both components are set equal to $0$ if the corresponding term does not appear in the query/document. Given these representations, the ranking function $s(q,d)$ can be written as
% \begin{equation}
%     s(q,d) = \langle \phi(q), \psi(d) \rangle = \sum_{i=1}^{|\mathcal{V}|} \phi_i(q)\psi_i(d),
% \end{equation}
% where $\langle \cdot, \cdot \rangle$ is the inner product in $\mathbb{R}^{|\mathcal{V}|}$. The specific form of the term-document score $s(q,t)$ is designed by hand according to some axioms or heuristics.
%
The aggregation function $f$ over these representations is an explicit formula taking into account the components of the query and document representations, i.e., the in-query and in-document term frequencies, together with other document normalisation operations.
These representations are referred to as \textit{sparse representations}, since most of their components are equal to $0$ because they correspond to tokens not appearing in the query/document.
Sparse representations can be trivially computed and efficiently stored in specialised data structures called \textit{inverted indexes}, which represent the backbone of commercial Web search engine~\cite{barla:2015}; see~\cite{manning:2008,buttcher:2010,fntir} for more details on inverted indexes and classical IR ranking models.

\subsection{LTR Features}\label{ssec:ltr}

With the advent of the Web, new sources of relevance information about the documents have been made available. The importance of a Web page, e.g., PageRank, additional document statistics, e.g., term frequencies in the title or anchors text, and search engine interactions, e.g., clicks, can be exploited as relevance signals. Moreover, collaborative and social platforms such as Wikipedia, Twitter and Facebook represent new sources of relevance signals. These relevance signals have been exploited to build richer query and document representations in LTR.
The relevance signals extracted from queries and/or documents are called \textit{features}. There are various classes of these features~\cite{macdonald:2012,bendersky2011quality}, such as:
\begin{itemize}
    \item \textit{query-only} features, i.e., components of $\phi(q)$: query features with the same value for each document, such as query type, query length, and query performance predictors;
    \item \textit{query-independent} features, i.e., components of $\psi(d)$: document features with the same value for each query, such as importance score, URL length, and spam score;
    \item \textit{query-dependent} features, i.e., components of $\eta(q,d)$: document features that depend on the query, such as different term weighting models on different fields.
\end{itemize}
In LTR, the representation functions are hand-crafted: exploiting the relevance signals from heterogeneous information sources, the different components of query and document representations are computed with feature-specific algorithms. Hence, the representations $\phi(q)$, $\psi(d)$, and $\eta(q,d)$ are elements of vector spaces over $\mathbb{R}$, but whose dimensions depend on the number of hand-crafted query-only, query-independent, and query-dependent features, respectively. Moreover the different components of these vectors are heterogeneous, and do not carry any specific semantic meaning. Using these representations, in LTR the aggregation function $f$ is machine-learned, for example using logistic regression~\cite{10.5555/188490.188560}, gradient-boosted regression trees~\cite{burges2010ranknet} or neural networks~\cite{ranknet}; see~\cite{Liu09ftir} for a detailed survey. 

\subsection{Word Embeddings}\label{ssec:wordemb}

Both BOW encodings and LTR features are widely adopted in commercial search engines, but they suffer from several limitations.
On the one hand, semantically-related terms end up having completely different BOW encodings. Although the two terms \texttt{catalogue} and \texttt{directory} can be considered synonyms, their BOW encodings are completely different, with the single $1$ appearing in different components. Similarly, two different documents on a same topic can end up having two unrelated BOW encodings. On the other hand, LTR features create text representations by hand via feature engineering, with heterogeneous components and no explicit concept of similarity.

In the 1950s, many linguists formulated the \textit{distributional hypothesis}: words that occur in the same contexts tend to have similar meanings~\cite{harris}. According to this hypothesis, the meaning of words can be inferred by their usage together with other words in existing texts. Hence, by leveraging the large text collections available, it is possible to learn useful representations of terms, and devise new methods to use these representations to build up more complex representations for queries and documents.

%To overcome these limitations, distributional semantics aims at quantifying the semantic similarity between words leveraging the large document collections available. This quantification is based on the ``Distributional Hypothesis'': words that occur in the same contexts tend to have similar meanings~\cite{harris}. In distributional semantics, 
These learned representations are vectors in $\mathbb{R}^n$, with $n \ll |\mathcal{V}|$, called \textit{distributional representations} or \textit{word embeddings}. The number of dimensions $n$ ranges approximatively from 50 to 1000 components, instead of the vocabulary size $|\mathcal{V}|$. Moreover, the components of word embeddings are rarely $0$: they are real numbers, and can also have negative values. Hence, word embeddings are also referred to as \textit{dense representations}.
%the tokens in a query/document are represented by vectors in $\mathbb{R}^n$, with $n \ll |\mathcal{V}|$, called \textit{distributional representations}. 
Among the different techniques to compute these representations, there are algorithms to compute global representations of the words, i.e., a single fixed embedding for each term in the vocabulary, called \textit{static word embeddings}, and algorithms to compute local representations of the terms, which depend on the other tokens used together with a given term, i.e., its context, called \textit{contextualised word embeddings}. Static word embeddings used in neural IR are learned from real-world text with no explicit training labels: the text itself is used in a self-supervised fashion to compute word representations. There are different kinds of static word embeddings, for different languages, such as \wordtovec~\cite{word2vec}, \fasttext~\cite{fasttext} and \glove~\cite{glove}.
Static word embeddings map terms with multiple senses into an average or most common sense representation based on the training data used to compute the vectors; each term in the vocabulary is associated with a single vector. Contextualised word embeddings map tokens used in a particular context to a specific vector; each term in the vocabulary is associated with a different vector every time it appears in a document, depending on the surrounding tokens. The most popular contextualised word embeddings are learned with deep neural networks such as the Bidirectional Encoder Representations from Transformers (\bert)~\cite{bert}, the Robustly Optimized \bert\ Approach (\roberta)~\cite{roberta}, and the Generative Pre-Training models (\gpt)~\cite{gpt}.

In neural IR, word embeddings are used to compute the representation functions $\phi$, $\psi$ and $\eta$, and the aggregation function $f$ through (deep) neural networks. 
Depending on the assumptions over the representation functions, the neural ranking models can be classified in interaction-focused models and representation-focused models. 
In interaction-focused models, the query-document representation function $\eta(q,d)$, taking into account the interaction between the query and document contents, is explicitly constructed and used as input to a deep neural network, or it is implicitly generated and directly used by a deep neural network.
In representation-focused models, the query-document representation function $\eta(q,d)$ is not present; query and document representations $\phi(q)$ and $\psi(d)$ are computed independently by deep neural networks
%The research in neural IR has been strongly influenced by two milestones: the introduction of a practical algorithms to compute word embeddings~\cite{word2vec} and the introduction of the \textit{transformer} neural network architecture to compute contextualised word embeddings~\cite{transformer}.
%In the following, we discuss the main representation-focused architectures (Section~\ref{sec:repr}) and interaction-focused architectures (Section~\ref{sec:aggr}). 
In the following, we discuss the main interaction-focused models for ad-hoc ranking (Section~\ref{sec:interaction}) and representation-focused models for queries and documents (Section~\ref{sec:representation}).

\section{Interaction-focused Systems}\label{sec:interaction}

The interaction-focused systems used in neural IR model the word and $n$-gram relationships across a query and a document using deep neural networks. These systems receive as input both a query $q$ and a document $d$, and output a query-document representation $\eta(q,d)$.  Among others, two neural network architectures have been investigated to build a representation of these relationships: convolutional neural networks and transformers. Convolutional neural networks represent one of the first approaches in building joint representations of queries and documents, as discussed in Section~\ref{ssec:cnn}. Transformers represent the major turning point in neural IR, as their application to textual inputs gave birth to pre-trained language models, presented in Section~\ref{ssec:plm}. 
In neural IR, pre-trained language models are used to compute query-document representations, and the two main transformer models used for this task are \bert\ and \tfive\, illustrated in Sections~\ref{ssec:bert}~and~\ref{ssec:t5}, respectively.
Section~\ref{ssec:finetuning1} describes how pre-trained language models are fine-tuned to compute effective query-document representations, and Section~\ref{ssec:passaging} briefly discusses how pre-trained language models can deal with long documents.

\subsection{Convolutional Neural Networks}\label{ssec:cnn}

\begin{figure}[htb!]
    \ifigure{.90}{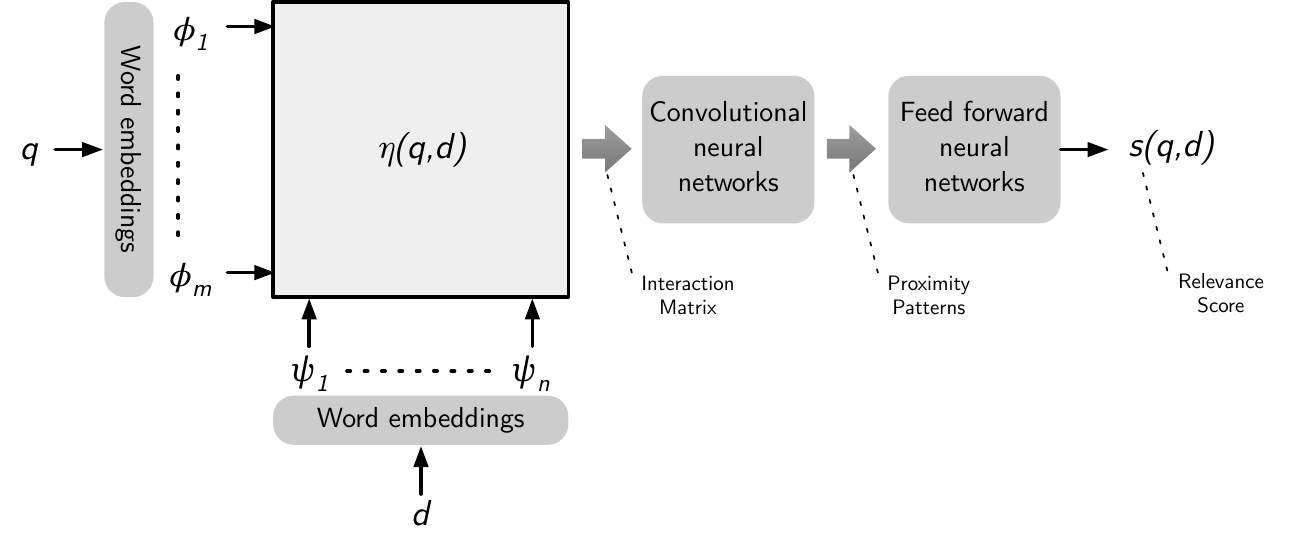}
    \caption{Scheme of an interaction-focused model based on convolutional neural networks.}\label{fig:cnn}
\end{figure}

A convolutional neural network is a family of neural networks designed to capture local patterns in structured inputs, such as images and texts~\cite{cnn}. The core component of a convolutional neural network is the \textit{convolution} layer, used in conjunction with feed forward and pooling layers. A convolutional layer can be seen as a small linear filter, sliding over the input and looking for proximity patterns.
Several neural models employ convolutional neural networks over the interactions between queries and documents to produce relevance scores. Typically, in these models, the word embeddings of the query and document tokens are aggregated into an \textit{interaction matrix}, on top of which convolutional neural networks are used to learn hierarchical \textit{proximity patterns} such as unigrams, bigrams and so on. Then, the final top-level proximity patterns are fed into a feed forward neural network to produce the relevance score $s(q,d)$ between the query $q$ and the document $d$, as illustrated in Figure~\ref{fig:cnn}.

The query $q$ and the document $d$ are tokenised into $m$ and $n$ tokens, respectively, and each token is mapped to a corresponding static word embedding. The interaction matrix $\eta(q,d) \in \mathbb{R}^{m \times n}$ is composed of the cosine similarities between a query token embedding and a document token embedding.

One of the first neural models leveraging the interaction matrix is the Deep Relevance Matching Model (\drmm)~\cite{drrm}). In \drmm, the cosine similarities of every query token w.r.t. the document tokens are converted into a discrete distribution using \textit{hard bucketing}, i.e., into a query token histogram. Then the histogram of each query token is provided as input to a feed forward neural network to compute the final query token-document relevance score. These relevance scores are then aggregated through an IDF-based weighted sum across the different query terms.
Instead of using hard bucketing, the Kernel-based Neural Ranking Model (\knrm)~\cite{knrm} proposes to use Gaussian kernels to smoothly distribute the contribution of each cosine similarity across different buckets before providing the histograms with \textit{soft bucketing} to the feed forward neural networks. 

Both \drmm\ and \knrm\ exploit the interaction matrix, but they do not incorporate any convolutional layer. In the Convolutional \knrm\ model (\convknrm)~\cite{convknrm}, the query and document embeddings are first independently processed through $k$ convolutional neural networks, to build unigam, bigram, up to $k$-gram embeddings. These convolutions allow to build word embeddings taking into account multiple close words at the same time. Then, $k^2$ cosine similarity matrices are built, between each combination of query and document $n$-gram embeddings, and these matrices are processed with \knrm. In the Position-Aware Convolutional Recurrent Relevant model (\pacrr)~\cite{pacrr}, the interaction matrix is processed through several convolutional and pooling layers to take into account words proximity. Convolutional layers are used also in other similar neural models~\cite{copacrr,deeprank,arc,matchpyramid,hint}.

%\subsection{Transformers}
\subsection{Pre-trained Language Models}\label{ssec:plm}

Static word embeddings map words with multiple senses into an average or most common-sense representation based on the training data used to compute the vectors. The vector of a word does not change  with the other words used in a sentence around it. 
The \textit{transformer} is a neural network designed to explicitly take into account the context of arbitrary long sequences of text, thanks to a special neural layer called \textit{self-attention}, used in conjunction with feed forward and linear layers. The self-attention layer maps input sequences to output sequences of the same length. When computing the $i$-th output element, the layer can access all the $n$ input elements (bidirectional self-attention) or only the first $i$ input elements (causal self-attention). A self-attention layer allows the network to take into account the relationships among different elements in the same input. When the input elements are tokens of a given text, a self-attention layer computes token representations that take into account their context, i.e., the surrounding words. In doing so, the transformer computes \textit{contextualised word embeddings}, where the representation of each input token is conditioned by the whole input text.

Transformers have been successfully applied to different natual language processing tasks, such as machine translation, summarisation, question answering and so on. All these tasks are special instances of a more general task, i.e., transforming an input text sequence to some output text sequence.
%The \textit{encoder-decoder} model, also known as \textit{sequence-to-sequence} model, has been designed to address this general task. 
The \textit{sequence-to-sequence} model has been designed to address this general task. 
The sequence-to-sequence neural network is composed of two parts: an \textit{encoder} model, which receives an input sequence and builds a contextualised representation of each input element, and a \textit{decoder} model, which uses these contextualised representations to generate a task-specific output sequence. Both models are composed of several stacked transformers. The transformers in the encoder employ bidirectional self-attention layers on the input sequence or the output sequence of the previous transformer. The transformers in the decoder employ causal self-attention on the previous decoder transformer's output and bidirectional cross-attention of the output of the final encoder transformer's output.
 
In neural IR, two specific instances of the sequence-to-sequence models have been studied: encoder-only models and encoder-decoder models. 
\textit{Encoder-only models} receive as input all the tokens of a given input sentence, and they compute an output contextualised word embedding for each token in the input sentence. Representatives of this family of models include \bert~\cite{bert}, \roberta~\cite{roberta}, and \distilbert~\cite{distilbert}.
\textit{Encoder-decoder models} generate new output sentences depending on the given input sentence. The encoder model receives as input all the tokens of a given sequence and builds a contextualised representation, and the decoder model sequentially accesses these embeddings to generate new output tokens, one token at a time. Representatives of this family of models include \bart~\cite{bart} and \tfive~\cite{t5}.

%\subsection{Pre-trained Language Models}

Sequence-to-sequence models can be trained as language models, by projecting with a linear layer every output embedding to a given vocabulary and computing the tokens probabilities with a \softmax\ operation.
The \softmax\ operation is a function $\sigma: \mathbb{R}^k \to [0,1]^k$ that takes as input $k > 1$ real values $z_1, z_2, \ldots, z_k$ and transforms each input $z_i$ as follows:
\begin{equation}\label{eq:softmax}
    \sigma(z_i) = \frac{e^{z_i}}{\sum_{j=1}^k e^{z_j}}
\end{equation}
\looseness -1 The \softmax\ operation normalises the input values into a probability distribution. In the context of deep learning, the inputs of a \softmax\ operation are usually called \textit{logits}, and they represent the raw predictions generated by a multi-class classification model, turned into a probability distribution over the classes by the \softmax\ operation.

Depending on the training objective, a sequence-to-sequence model can be trained as a \textit{masked language model} (MLM), as for \bert, or a \textit{casual language model} (CLM), as for \tfive. MLM training focuses on learning to predict missing tokens in a sequence given the surrounding tokens; CLM training focuses on predicting the next token in an output sequence given the preceding tokens in the input sequence. In both cases, it is commonplace to train these models using massive text data to obtain \textit{pre-trained language models}. In doing so, we allow the model to learn general-purpose knowledge about a language that can be adapted afterwards to a more specific \textit{downstream task}. In this \textit{transfer learning} approach, a pre-trained language model is used as initial model to fine-tune it on a domain-specific, smaller training dataset for the downstream target task. In other words, \textit{fine-tuning} is the procedure to update the parameters of a pre-trained language model for the domain data and target task.

\begin{figure}[htb!]
    \ifigure{.90}{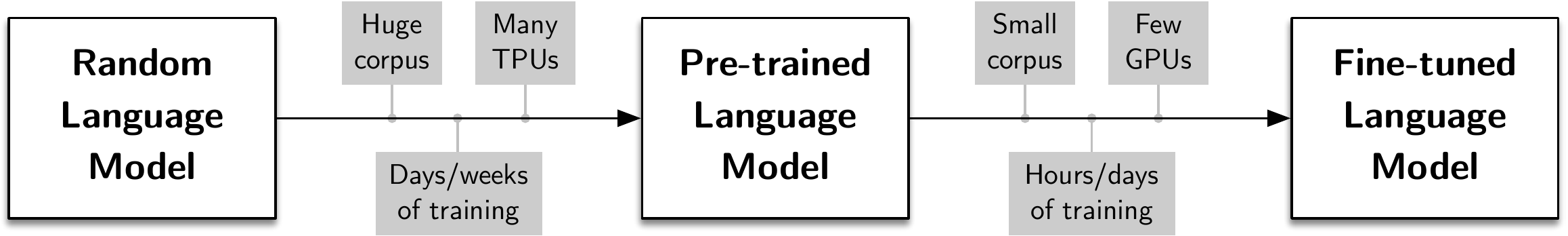}
    \caption{Transfer learning of a pre-trained language model to a fine-tuned language model.}\label{fig:plm}
\end{figure}

As illustrated in Figure~\ref{fig:plm}, pre-training typically requires a huge general-purpose training corpus, such as Wikipedia or Common Crawl web pages, expensive computation resources and long training times, spanning several days or weeks. On the other side, fine-tuning requires a small domain-specific corpus focused on the downstream task, affordable computational resources and few hours or days of additional training. Special cases of fine-tuning are \textit{few-shot learning}, where the domain-specific corpus is composed of a very limited number of training data, and \textit{zero-shot learning}, where a pre-trained language model is used on a downstream task that it was not fine-tuned on.

In neural IR, the interaction-focused systems that use pre-trained language models are called \textit{cross-encoder} models, as they receive as input a pair $(q,d)$ of query and document texts. Depending on the type of sequence-to-sequence model, different cross-encoders are fine-tuned in different ways, but, in general, they aim at computing a relevance score $s(q,d)$ to rank documents w.r.t. a given query. In the following, we illustrate the most common cross-encoders leveraging both encoder-only models (Sec.~\ref{ssec:bert}) and encoder-decoder models (Sec.~\ref{ssec:t5}).
%\mynote{Cascading can go here? No sure...}
%\subsection{BERT for Ranking}

\subsection{Ranking with Encoder-only Models}\label{ssec:bert}

The most widely adopted transformer architecture in neural IR is \bert, an encoder-only model. 
%BERT is an encoder-only models, composed of a stack of encoders. 
Its input text is tokenised using the WordPiece sub-word tokeniser~\cite{wordpiece}. The vocabulary $\mathcal{V}$ of this tokeniser is composed of $30,522$ terms, where the uncommon/rare words, e.g., \texttt{goldfish}, are splitted up in sub-words, e.g., \texttt{gold\#\#} and \texttt{\#\#fish}. The first input token of \bert\ is always the special \cls\ token, that stands for ``classification''. \bert\ accepts as input other special tokens, such as \sep, that denotes the end of a text provided as input or to separate two different texts provided as a single input. \bert\ accepts as input at most $512$ tokens, and produces an output embedding in $\mathbb{R}^\ell$ for each input token. The most commonly adopted \bert\ version is \bert{\sf\ base}, which stacks $12$~transformer layers, and whose output representation space has $\ell = 768$~dimensions.
%Every input token, including the special tokens, is then replaced with a combination of its static word embedding, a positional embedding, encoding the position of the token in the input text, and a sentence embedding, encoding different sentences in the same input text.

\citet{bertir}~and~\citet{cedr} illustrated how to fine-tune \bert\ as a cross-encoder\footnote{\citet{bertir} call it \textit{monoBERT}, and \citet{cedr} call it \textit{vanilla BERT}.}, in two slightly different ways.
Given a query-document pair, both texts are tokenised into token sequences $q_1, \ldots, q_m$ and $d_1, \ldots, d_n$. Then, the tokens are concatenated with \bert\ special tokens to form the following input configuration:
\begin{equation*}
    \cls\;q_1 \cdots q_m\;\sep\;d_1 \cdots d_n\;\sep
\end{equation*}
that will be used as \bert\ input. In doing so, the self-attention layers in the \bert\ encoders are able to take into account the semantic interactions among the query tokens and the document tokens. The output embedding $\eta_{\cls} \in \mathbb{R}^\ell$, corresponding to the input \cls\ token, serves as a contextual representation of the query-document pair as a whole. 

\begin{figure}[htb!]
    \ifigure{.50}{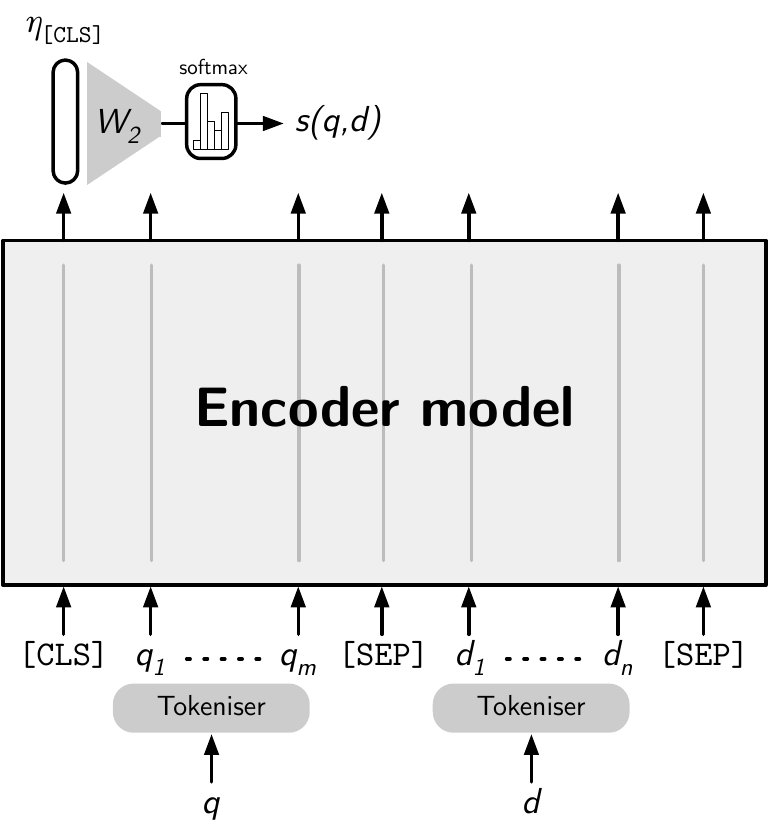}
    \caption{\bert\ classification model for ad-hoc ranking.}\label{fig:bert}
\end{figure}

\citet{monobert} fine-tune \bert\ on a binary classification task to compute the query-document relevance score, as illustrated in Figure~\ref{fig:bert}.
%classification layer, i.e., a linear projection layer acting as a binary classifier between the ``relevant'' and ``non-relevant'' classes.
To produce the relevance score $s(q,d)$, the query and the document are processed by \bert\ to generate the output embedding $\eta_{\cls} \in \mathbb{R}^\ell$, that is multiplied by a learned set of classification weights $W_2 \in \mathbb{R}^{2 \times \ell}$ to produce two real scores $z_0$ and $z_1$, and then through a \softmax\ operation to transform the scores into a probability distribution $p_0$ and $p_1$ over the non-relevant and relevant classes. The probability corresponding to the relevant class, conventionally assigned to label $1$, i.e., $p_1$, is the final relevance score.

\citet{cedr} fine-tune \bert\ by projecting the output embedding $\eta_{\cls} \in \mathbb{R}^\ell$ through the learned matrix $W_1 \in \mathbb{R}^{1 \times \ell}$ into a single real value $z$, that represents the final relevance score.

% \begin{equation}
%      \begin{split}
%         \eta_{\cls} &= {\bert}(q,d)\\
%         [z_0, z_1] &= W_C y_{\cls} \\
%         [p_0, p_1] &= \text{softmax}([z_0,z_1])\\
%         s(q,d) &= p_1
%     \end{split}
% \end{equation}
% More formally, to produce the relevance score $s(q,d)$, the query and the document are processed by \bert\ to generate the output embedding $y_{\cls} \in \mathbb{R}^\ell$, that is multiplied by a learned set of projection weights $W$. Depending on the loss function adopted to fine-tune the model, the linear layer using the weights $W$ can output two values $z_0$ and $z_1$, over the ``non-relevant'' and ``relevant'' classes, i.e., $W = W_2 \in \mathbb{R}^{2 \times \ell}$, or just a single value $z$, i.e., $W = W_1 \in \mathbb{R}^{1 \times \ell}$. The relevance score for the query-document pair is conventionally $z_1$ in the first case, or simply $z$ in the second case:
\begin{equation}\label{eq:bert}
 \begin{alignedat}{2}
      \eta_{\cls} &= {\bert}(q,d)\\
     [z_0, z_1] &= W_2  \eta_{\cls} \quad \text{or} \quad z = W_1 \eta_{\cls}\\
     [p_0, p_1] &= {\softmax}([z_0,z_1])\\
     s(q,d) &= p_1 \quad \text{or} \quad s(q,d) = z\\
 \end{alignedat}
\end{equation}

In Section~\ref{ssec:finetuning1} we illustrate how a \bert-based cross-encoder is typically fine-tuned for ad-hoc ranking.

% This cross-encoder is fine-tuned using a binary cross entropy loss. Given a training triple $(q, d^+, d^-)$ with a query $q$, a relevant document $d^+$ and non-relevant document $d^-$, this loss is computed as:
% \begin{equation}\label{eq:bce}
%     \mathcal{L}(q, d^+, d^-) = -\log\big(s(q,d^+)\big) -\log\big(1-s(q,d^-)\big)
% \end{equation}
%\citet{monobert} refer to this cross-encoder as \textit{monoBERT}.

\subsection{Ranking with Encoder-decoder Models}\label{ssec:t5}

Instead of using an encoder-only transformer model to compute the latent representation of a query-document pair and to convert it into a relevance score, it is possible also to use an encoder-decoder model~\cite{t5} with \textit{prompt learning}, by converting the relevance score computation task into a cloze test, i.e., a fill-in-the-blank problem. Prompting has been successfully adopted in article summarisation tasks~\cite{Radford2019LanguageMA} and knowledge base completion tasks~\cite{petroni-etal-2019-language}.

In prompt learning, the input texts are reshaped as a natural language template, and the downstream task is reshaped as a cloze-like task. For example, in topic classification, assuming we need to classify the sentence \textit{text} into two classes $c_0$ and $c_1$, the input template can be:
\begin{equation*}
    \texttt{Input:}\;\;\textit{text}\;\;\;\texttt{Class:}\;\;\out
\end{equation*}
Among the vocabulary terms, two \textit{label terms} $w_0$ and $w_1$ are selected to correspond to the classes $c_0$ and $c_1$, respectively. The probability to assign the input \textit{text} to a class can be transferred into the probability that the input token \out\ is assigned to the corresponding label token:
\begin{equation*}
\begin{split}
    p(c_0|\textit{text}) = p(\out = w_0 | \texttt{Input:}\;\;\textit{text}\;\;\;\texttt{Class:}\;\;\out)\\
    p(c_1|\textit{text}) = p(\out = w_1 | \texttt{Input:}\;\;\textit{text}\;\;\;\texttt{Class:}\;\;\out)
\end{split}
\end{equation*}

\begin{figure}[htb!]
    \ifigure{.60}{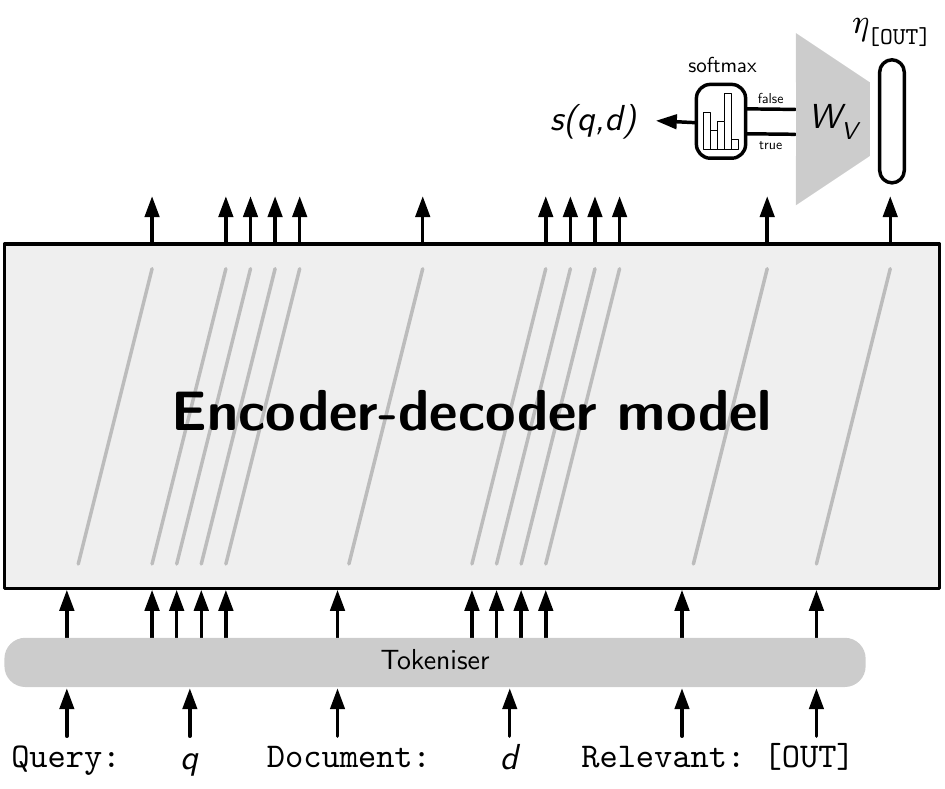}
    \caption{\tfive\ model for ad-hoc ranking.}\label{fig:t5}
\end{figure}

\citet{monot5} proposed a prompt learning approach for relevance ranking using a \tfive\ model\footnote{\citet{monot5} call it \textit{monoT5}.}, as illustrated in Figure~\ref{fig:t5}. The query and the document texts $q$ and $d$ are concatenated to form the following input template:
\begin{equation*}
    \texttt{Query:}\;\;q\;\;\texttt{Document:}\;\;d\;\;\texttt{Relevant:}\;\;\out
\end{equation*}
An encoder-decoder model is fine-tuned with a downstream task taking as input this input configuration, and generating an output sequence whose last token is equal to {\tt True} or {\tt False}, depending on whether the document $d$ is relevant or non-relevant to the query $q$.

The query-document relevance score is computed by normalising only the {\tt False} and {\tt True} output probabilities, computed over the whole vocabulary, with a \softmax\ operation.

To produce the relevance score $s(q,d)$, the query and the document are processed by \tfive\ to generate the output embedding $\eta_{\out} \in \mathbb{R}^\ell$, that is projected by a learned set of classification weights $W_V \in \mathbb{R}^{|\mathcal{V}| \times \ell}$ over the vocabulary $\mathcal{V}$. The outputs $z_{\texttt{False}}$ and $z_{\texttt{True}}$ corresponding to the {\tt False} and {\tt True} terms, respectively, are transformed into a probability distribution with a \softmax\ operation, to yield the required predictions $p_{\texttt{False}}$ and $p_{\texttt{True}}$ over the ``non-relevant'' and ``relevant'' classes. The prediction corresponding to the relevant class, i.e., $p_{\texttt{True}}$, is the final relevance score.

\begin{equation}\label{eq:t5}
    \begin{split}
        \eta_{\out} &= {\tfive}(q,d)\\
        [\ldots, z_{\texttt{False}}, \ldots, z_{\texttt{True}}, \ldots]^T & = W_V \eta_{\out}\\
        [p_{\texttt{False}}, p_{\texttt{True}}] &= {\softmax}([z_{\texttt{False}}, z_{\texttt{True}}])\\
        s(q,d) &= p_{\texttt{True}}\\
    \end{split}
\end{equation}

%\citet{monot5} refer to this cross-encoder as \textit{monoT5}.
In the next section we discuss how a \tfive-based cross-encoder is typically fine-tuned for ad-hoc ranking.

\subsection{Fine-tuning Interaction-focused Systems}\label{ssec:finetuning1}

As discussed in Section~\ref{ssec:plm}, the pre-trained language models adopted in IR require a fine-tuning of the model on a specific downstream task. Given an input query-document pair $(q,d)$ a neural IR model $\mathcal{M}(\theta)$, parametrised by $\theta$, computes $s_\theta(q,d) \in \mathbb{R}$, the score of the document $d$ w.r.t. the query $q$. We are supposed to predict $y \in \{+, -\}$ from $(q,d) \in \mathcal{Q} \times \mathcal{D}$, where $-$ stands for non-relevant and $+$ for relevant. 
%Given a query $q$, it is commonplace to denote with $d^+$ a relevant document and with $d^-$ a non-relevant document.
%We perform the prediction by assuming a joint distribution $p(x,y)$ from which we can sample correct pairs $(y,x)$ and learning a score function $s_\theta(q,d)$ such that it assigns a high score to a relevant document and a low score to a non-relevant document.

This problem can be framed as a binary classification problem. We perform the classification by assuming a joint distribution $p$ over $\{+,-\} \times \mathcal{Q} \times \mathcal{D}$ from which we can sample correct pairs $(+, q,d) \equiv (q,d^+)$ and $(-, q,d) \equiv (q, d^-)$.
Using sampled correct pairs we learn a score function $s_\theta(q,d)$ 
as an instance of a metric learning problem~\cite{metric:learning}:
the score function must assign a high score to a relevant document and a low score to a non-relevant document, as in Eq.~\eqref{eq:bert} and Eq.~\eqref{eq:t5}.
Then we find $\theta^*$ that minimises the (binary) cross entropy $l_{CE}$ between the conditional probability $p(y|q,d)$ and the model probability $p_\theta(y|q,d)$:
\begin{equation}
    \theta^* = \arg\min_\theta \mathbb{E}\Big[l_{CE}(y,q,d)\Big]\\
\end{equation}
where the expectation is computed over $(y,q,d) \sim p$, and the cross entropy
%for a relevant document $d^+$ and a non-relevant document $d^-$ for the same query $q$
is computed as:
\begin{equation}
    l_{CE}\big(y,(q,d)\big) = 
    \begin{cases}
        -\log\big(s_\theta(q,d)\big)\;&\text{if}\;y=+\\
        -\log\big(1 - s_\theta(q,d)\big) \;&\text{if}\;y=-\\
    \end{cases}\\
\end{equation}
Typically, a dataset $\mathcal{T}$ available for fine-tuning pre-trained language models for relevance scoring is composed of a list of triples $(q, d^+, d^-)$, where $q$ is a query, $d^+$ is a relevant document for the query, and $d^-$ is a non-relevant document for the query. In this case, the expected cross entropy is approximated by the sum of the cross entropies computed for each triple:
\begin{equation}
    %l_{CE}(q, d^+, d^-) = -\log\big(s_\theta(q,d^+)\big) -\log\big(1 - s_\theta(q,d^-)\big)\\
    \mathbb{E}\Big[l_{CE}\big(y,(q,d)\big)\Big] \approx \frac{1}{2|\mathcal{T}|}\sum_{(q, d^+, d^-) \in \mathcal{T}} \Big(-\log\big(s_\theta(q,d^+)\big) -\log\big(1 - s_\theta(q,d^-)\big)\Big)
\end{equation}
In doing so, we do not take into account documents in the collection that are not explicitly labeled as relevant or non-relevant. This approach is limited to take into account positive and negative triples in a pairwise independent fashion. In Section~\ref{ssec:finetuning2} we will discuss a different fine-tuning approach commonly used for representation-focused systems, taking into account multiple non-relevant documents per relevant document.

\subsection{Dealing with long texts}\label{ssec:passaging}

\bert\ and \tfive\ models have an input size limited to 512 tokens, including the special ones. When dealing with long documents, that cannot be fed completely into a transformer model, we need to split them into smaller texts in a procedure referred to as \textit{passaging}. \citet{10.1145/3331184.3331303} propose to split a long document into overlapping shorter passages, to be processed independently together with the same query by a cross-encoder. During training, if a long document is relevant, all its passages are relevant, and vice-versa. Then, the relevance scores for each composing passage are aggregated back to a single score for the long document. In this scenario, the common aggregation functions are \firstp, i.e., the document score is the score of the first passage, \maxp, i.e., the document score is the highest score across all passages, and \sump, i.e., the document score is the sum of the scores of its passages. Alternatively, \citet{li2020parade} generate the \cls\ output embedding for each passage to compute a query-passage representation for each passage. Then, the different passage embeddings are aggregated together to compute a final relevance score for the whole document using feed forward neural networks, convolutional neural networks or simple transformer architectures.

\section{Representation-focused Systems}\label{sec:representation}

The representation-focused systems build up independent query and document representations, in such a way that document representations can be pre-computed and stored in advance. During query processing, only the query representation is computed, and the top documents are searched through the stored document representations.
In doing so, representation-based systems are able to identify the relevant documents among all documents in a collection, rather than just among a query-dependent sample; these systems represent a new type of retrieval approaches called \emph{dense retrieval} systems.
Thus far, two different families of representations have emerged in dense retrieval. In \textit{single representations} systems, queries and documents are represented with a single embedding, as discussed in Section~\ref{ssec:singlerep}. In \textit{multiple representations} systems, queries and/or documents are represented with more than a single embedding, as discussed in Section~\ref{ssec:multiplerep}. Section~\ref{ssec:finetuning2} discusses how representation-focused systems are fine-tuned, exploiting noise-contrastive estimation.

\subsection{Single Representations}\label{ssec:singlerep}

In interaction-focused systems discussed in Section~\ref{sec:interaction}, the query and document texts are concatenated together before processing with sequence-to-sequence models, yielding rich interactions between the query context and the document context, as every word in the document can attend to every word in the query, and vice-versa. At query processing time, every document must be concatenated with the query and must be processed with a forward pass of the whole sequence-to-sequence model. Even if some techniques such as the pre-computation of some internal representations have been proposed~\cite{prettr}, interaction-focused systems cannot scale to a large amount of documents. In fact, on a standard CPU, the processing of a query over the whole document collection exploiting an inverted index requires a few milliseconds~\cite{fntir}, while computing the relevance score of a single query-document pair with a transformer model may requires a few seconds~\cite{cedr}.

\begin{figure}[ht!]
    \ifigure{.75}{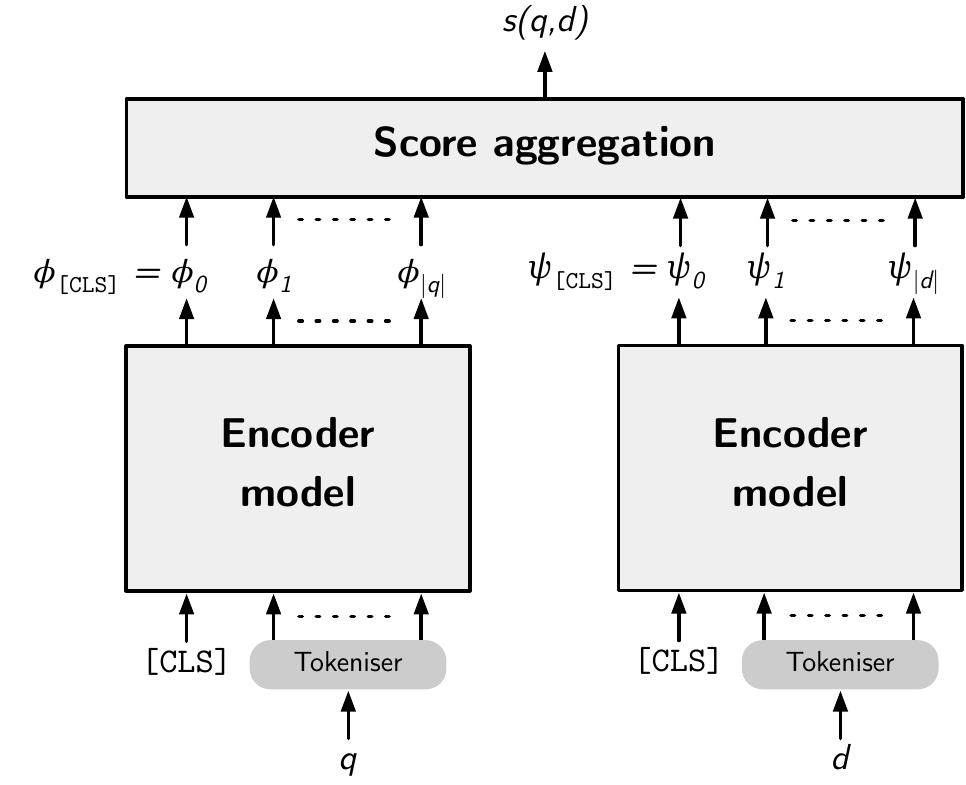}
    \caption{Representation-focused system.}\label{fig:bienc}
\end{figure}

Instead of leveraging sequence-to-sequence models to compute a semantically richer but computationally expensive interaction representation $\eta(q,d)$, representation-focused systems employ encoder-only models to independently compute query representations $\phi(q)$ and document representations $\psi(d)$ in the same latent vector space~\cite{urbanek-etal-2019-learning}, as illustrated in Figure~\ref{fig:bienc}. Next, the relevance score between the representations is computed via an aggregation function $f$ between these representations:
\begin{equation}
    \begin{split}
        \phi(q) &= [\phi_\cls, \phi_1, \ldots, \phi_{|q|}] = {\encoder}(q)\\
        \psi(d) &= [\psi_\cls, \psi_1, \ldots, \psi_{|d|}] = {\encoder}(d)\\
%        s(q,d) &= \langle \psi(q), \phi(d) \rangle = \psi(q) \cdot \phi(d).
        %s(q,d) &= \psi(q) \cdot \phi(d)
        %s(q,d) &= y^q_0 \cdot y^d_0
        s(q,d) &= f\big(\phi(q), \psi(d)\big)
    \end{split}
\end{equation}
In neural IR, the representation functions $\phi$ and $\psi$ are computed through fine-tuned encoder-only sequence-to-sequence models such as \bert. The same neural model is used to compute both the query and the document representations, so the model is also called \textit{dual encoder} or \textit{bi-encoder}~\cite{NIPS1993_288cc0ff}. A bi-encoder maps queries and documents in the same vector space $\mathbb{R}^\ell$, in such a way that the representations can be mathematically manipulated. Usually, the output embedding corresponding to the \cls\ token is assumed to be the representation of a given input text. Using these \textit{single representations}, the score aggregation function is the dot product:
\begin{equation}
    %s(q,d) = f\big(\phi(q), \psi(d)\big) = \phi_\cls \cdot \psi_\cls
    s(q,d) = \phi_\cls \cdot \psi_\cls
\end{equation}

Different single-representation systems have been proposed: \dpr~\cite{dpr}, \ance~\cite{ance}, and \star~\cite{star} being the most widely adopted. The main difference among these systems is how the fine-tuning of the \bert\ model is carried out, as discussed in Section~\ref{ssec:finetuning2}.

\subsection{Multiple Representations}\label{ssec:multiplerep}

Up so far, we considered representation-focused systems in which queries and documents are represented through a single embedding in the latent vector space.
%Although this single embedding can be obtained by different aggregation schemes over the output embeddings of an encoder-only model, neural IR systems such as DPR~\cite{dpr}, ANCE~\cite{ance} and STAR~\cite{star} use the first output embedding, i.e., the output embedding corresponding to the \cls input special token. 
This single representation is assumed to incorporate the meaning of an entire text within that single embedding.

In contrast, \textit{multiple representation} systems such as poly-encoders~\cite{polyencoder}, \mebert~\cite{mebert}, \colbert~\cite{colbert} and \coil~\cite{coil} exploit more than a single embedding to represent a given text,  which may allow a richer semantic representation of the content. 

Instead of using just the first output embedding $\psi_{\cls} = \psi_0$ to encode a document $d$, poly-encoders~\cite{polyencoder}  exploit the first $m$ output embeddings $\psi_0, \psi_1, \ldots, \psi_{m-1}$. A query $q$ is still represented with the single embedding  $\phi_{\cls} = \phi_0$, while we need to aggregate the $m$ output document embeddings into a single representation $\psi_*$ to compute the final relevance score using the dot product with the output query embedding. To do so, poly-encoders first compute the $m$ similarities $s_0, \ldots, s_{m-1}$ between the query embedding and the first $m$ document embedding using the dot product. These similarities are transformed into normalised weights $v_0, \ldots, v_{m-1}$ using a \softmax\ operation, and the weighted output embeddings are summed up to compute the final document embedding $\psi_*$ used to compute the relevance~score:
\begin{equation}\label{eq:polyenc}
    \begin{split}
        [\phi_0, \phi_1, \ldots] &= {\encoder}(q)\\
        [\psi_0, \psi_1, \ldots] &= {\encoder}(d)\\
        [s_0, s_1, \ldots, s_{m-1}] &= [\phi_0 \cdot \psi_0, \phi_0 \cdot \psi_1, \ldots, \phi_0 \cdot \psi_{m-1}]\\
        [v_0, v_1, \ldots, v_{m-1}] &= {\softmax}([s_0, s_1, \ldots, s_{m-1}])\\
        \psi_* &= \sum_{i=0}^{m-1} v_i \psi_i\\
        s(q,d) &= \phi_0 \cdot \psi_*\\
    \end{split}
\end{equation}

Similarly to poly-encoders, \mebert~\cite{mebert} exploits the first $m$ output embeddings to represent a document $d$ (including the \cls\ embedding), but uses a different strategy to compute the relevance score $s(q,d)$ w.r.t. a query $q$. \mebert\ computes the similarity between the query embedding and the first $m$ document embedding using the dot product, and the maximum similarity, also called \textit{maximum inner product}, represents the relevance score:
\begin{equation}\label{eq:maxsim}
    \begin{split}
        [\phi_0, \phi_1, \ldots] &= {\encoder}(q)\\
        [\psi_0, \psi_1, \ldots] &= {\encoder}(d)\\
        s(q,d) &= \max_{i=0,\ldots,m-1} \phi_0 \cdot \psi_i\\
    \end{split}
\end{equation}
This relevance scoring function, called \textit{max similarity} or \textit{maxsim}, allows us to exploit efficient implementations of maximum inner product search systems, discussed in Section~\ref{sec:dense}. On the contrary, the relevance scoring function in Eq.~\eqref{eq:polyenc}, based on a \softmax\ operation, does not permit to decompose the relevance scoring to a maximum computation over dot products.

Differently from poly-encoders and \mebert, \colbert~\cite{colbert} does not limit to $m$ the number of embeddings used to represent a document. Instead, it uses all the $1+|d|$ output embeddings to represent a document, i.e., one output embedding per document token, including the \cls\ special token. Moreover, also a query $q$ is represented with multiple $1+|q|$ output embeddings, i.e., one output embedding per query token, including the \cls\ special token. As in other representation-focused systems, query token embeddings are computed at query processing time; queries may also be augmented with additional \emph{masked tokens} to provide  ``{\em a soft, differentiable mechanism for learning to expand queries with new terms or to re-weigh existing terms based on their importance for matching the query}''~\cite{colbert}. In current practice, queries are augmented up to $32$ query token embeddings. 
Without loss of generality, query and documents embeddings can be projected in a smaller latent vector space through a learned weight matrix $W \in \mathbb{R}^{\ell' \times \ell}$, with $\ell' < \ell$.

Since there are multiple query embeddings, \colbert\ exploits a modified version of the relevance scoring function in Eq~\eqref{eq:maxsim}, where every query embedding contributes to the final relevance score by summing up its maximum dot product value w.r.t. every document embeddings:

\begin{equation}\label{eq:maxsim2}
    \begin{split}
        [\phi_0, \phi_1, \ldots] &= {\encoder}(q)\\
        [\psi_0, \psi_1, \ldots] &= {\encoder}(d)\\
        s(q,d) &= \sum_{i = 0}^{|q|}\max_{j=0,\ldots,|d|} \phi_i \cdot \psi_j
    \end{split}
\end{equation}

\colbert's late interaction scoring in Eq.~\eqref{eq:maxsim2}, also called \textit{sum maxsim}, performs an all-to-all computation: each query embedding, including the masked tokens' embeddings, is dot-multiplied with every document embedding, and then the maximum computed dot products for each query embedding are summed up. In doing so, a query term can contribute to the final scoring by (maximally) matching a different lexical token. A different approach is proposed by the \coil\ system~\cite{coil}. In \coil, the query and document \cls\ embeddings are linearly projected with a learned matrix $W_C \in \mathbb{R}^{\ell \times \ell}$. The embeddings corresponding to normal query and document tokens are projected into a smaller vector space with dimension $\ell' < \ell$, using another learned matrix $W_T \in \mathbb{R}^{\ell' \times \ell}$. Typical values for $\ell'$ range from $8$ to $32$.

The query-document relevance score is the sum of two components. The first component is the dot product of the projected query and document \cls\ embeddings, and the second component is the sum of sub-components, one per query token. Each sub component is the maximum inner product between a query token and the document embeddings for the same token:
\begin{equation}\label{eq:maxsim3}
    \begin{split}
        [\phi_0, \phi_1, \ldots] &= {\encoder}(q)\\
        [\psi_0, \psi_1, \ldots] &= {\encoder}(d)\\
        [\phi'_0, \phi'_1, \ldots] &= [W_C\phi_0, W_T\phi_1, \ldots]\\
        [\psi'_0, \psi'_1, \ldots] &= [W_C\psi_0, W_T\psi_1, \ldots]\\
        s(q,d) &= \phi'_0 \cdot \psi'_0 + \sum_{t_i \in q}\max_{t_j \in d, \, t_j = t_i} \phi'_i \cdot \psi'_j
    \end{split}
\end{equation}
The \coil's scoring function, based on lexical matching between query and document tokens, allows us to pre-compute the projected document embeddings and, for each term in the vocabulary, to concatenate together the embeddings in the same document and in the whole collection, organising them in posting lists of embeddings, including a special posting list for the \cls\ token and its document embeddings. This organisation permits the efficient processing of posting lists at query time with optimised linear algebra libraries such as \blas~\cite{blas}. Note that the projected query embeddings are still computed at query processing time.

\subsection{Fine-tuning Representation-focused Systems}\label{ssec:finetuning2}

The fine-tuning of a bi-encoder  corresponds to learning an appropriate inner-product function suitable for the ad-hoc ranking task, i.e., for relevance scoring. 
%In the basic metric learning model, we are concerned with learning an appropriate distance function tuned to some given task. As discussed previously, 
As in Section~\ref{ssec:finetuning1}, we have a neural IR model $\mathcal{M}(\theta)$, parametrised by $\theta$, that computes a score $s_\theta(q,d)$ for a document $d$ w.r.t. a query $q$. We now frame the learning problem as a probability estimation problem. To this end, we turn the scoring function into a proper conditional distribution by using a \softmax\ operation:
\begin{equation}\label{eq:pdq}
    p_\theta(d|q) = \frac{\exp\big(s_\theta(q,d)\big)}{\sum_{d' \in \mathcal{D}} \exp\big(s_\theta(q,d')\big) }
\end{equation}
where $p_\theta(d|q)$ represents the posterior probability of the document being relevant given the query. We assume to have a joint distribution $p$ over $\mathcal{D} \times \mathcal{Q}$, and we want to find the parameters $\theta^*$ that minimise the cross entropy $l_{CE}$ between the actual probability $p(d|q)$ and the model probability $p_\theta(d|q)$:
\begin{equation}\label{eq:ce2}
    \theta^* = \arg\min_\theta \mathbb{E}\Big[l_{CE}(d,q)\Big] = \arg\min_\theta \mathbb{E}\Big[-\log\big(p_\theta(d|q)\big)\Big]
\end{equation}
where the expectation is computed over $(d,q) \sim p$. If the scoring function $s_\theta(q,d)$ is expressive enough, then, for some $\theta$, we have $p(d|q) = p_\theta(d|q)$.

The cross entropy loss in Eq.~\eqref{eq:ce2} is difficult to optimise, since the number of documents in $\mathcal{D}$ is large, and then the denominator in Eq.~\eqref{eq:pdq}, also known as \textit{partition function}, is expensive to compute. In \textit{noise contrastive estimation} we choose an artificial noise distribution $g$ over $\mathcal{D}$ of \textit{negative samples} and maximise the likelihood of $p_\theta(d|q)$ contrasting $g(d)$. Given $k \geq 2$ documents $\mathcal{D}_k = \{d_1, \ldots, d_k\}$, for each of them we define the following conditional distribution:
\begin{equation}
    \hat{p}_\theta(d_i|q, \mathcal{D}_k) = \frac{\exp\big(s_\theta(q,d_i)\big)}{\sum_{d' \in \mathcal{D}_k} \exp\big(s_\theta(q,d')\big) }
\end{equation}
which is significantly cheaper to compute that Eq.~\eqref{eq:pdq} if $k \ll |\mathcal{D}|$. Now, we want to find the parameters $\theta^\dagger$ that minimise the noise contrastive estimation loss $l_{NCE}$, defined as:
\begin{equation}\label{eq:nce}
    \theta^\dagger = \arg\min_\theta \mathbb{E}\Big[l_{NCE}(\mathcal{D}_k,q)\Big] = \arg\min_\theta \mathbb{E}\Big[-\log\big(\hat{p}_\theta(d_1|q, \mathcal{D}_k)\big)\Big]
\end{equation}
where the expectation is computed over $(d_1, q) \sim p$ and $d_i \sim g$ for $i=2, \ldots, k$.

The end goal of this fine-tuning is to learn a latent vector space for query and document representations where a query and its relevant document(s) are closer, w.r.t. the dot product, than the query and its non-relevant documents~\cite{dpr}; this fine-tuning approach is also called \textit{contrastive learning}~\cite{dssm}.

Negative samples are drawn from the noise distribution $g$ over $\mathcal{D}$. In the following, we list some negative sampling strategies adopted in neural IR.\bigskip

\begin{itemize}
    \item \textit{Random sampling}: any random document from the corpus is considered non-relevant, with equal probability, i.e., $q(d) = 1/|\mathcal{D}|$. Any number of negative documents can be sampled. Intuitively, it is reasonable to expect that a randomly-sampled document will obtain a relevance score definitely smaller than the relevance score of a positive document, with a corresponding loss value close to $0$. Negative documents with near zero loss contribute little to the training convergence to identify the parameters $\theta^\dagger$~\cite{NEURIPS2018_967990de,Katharopoulos2018NotAS}.

    \item \textit{In-batch sampling}: during training, the queries to compute the loss can be randomly aggregated into batches of size $b$, for faster training. For each query in a given batch, the positive passages for the other $b-1$ queries are considered as negative passages for the query~\cite{gillick-etal-2019-learning}. This sampling approach suffers from the same near zero loss problem as random sampling~\cite{ance}, but the sampling prodedure is faster.
    
    \item \textit{Hard negative sampling}: negative documents can be generated exploiting a classical or trained retrieval system. Each query is given as input to the retrieval system, the top documents retrieved, and the documents not corresponding to the positive ones are treated as negatives. Note that in this case we are assuming a conditional noise distribution $p(d|q, d_1)$, since we assume to know the relevant document $d_1$ for the query. In doing so, high-ranking documents are prioritised w.r.t. low-ranking documents, that do not impact on the user experience and do not contribute to the loss. The retrieval system used to mine the negative documents can exploit BM25 relevance model, as in \dpr~\cite{dpr}, the currently neural model under training, as in \ance~\cite{ance}, or another fine-tuned neural model, as in \star~\cite{star}.
\end{itemize}

% DPR: metrci learning problem, positive and negative passages, random, BM25 and gold negatives
% STAR: static vs. dynamic hard negatives, small theoretic justification
% ANCE: contrastive learning during training

% Poi abbiamo colbert, me-bert, polyencoders

\section{Retrieval Architectures and Vector Search}\label{sec:dense}

This section illustrates how the neural IR systems discussed so far are actually deployed in end-to-end systems. Section~\ref{ssec:architectures} discusses how cross-encoders and bi-encoders are deployed in ranking architectures. Since dense retrieval systems pre-computed the document embeddings, many actual systems focus on storing and searching through document embeddings. Section~\ref{ssec:search} formally defines the search problems in vector spaces and the embedding index, while Sections~\ref{ssec:lsh}, \ref{ssec:pq}, and \ref{ssec:hnsw} illustrate different solutions for efficiently storing and searching vectors. Section~\ref{ssec:optimisations} discusses some optimisations for embedding indexes specifically tailored for dense retrieval systems.
% \nic{In dense retrieval, passages are represented by real-valued vectors, \iadh{while the} query-document similarity is computed by deploying efficient nearest neighbour techniques over specialised indexes, such as \craig{those provided by the FAISS toolkit}~\cite{JDH17}}. %\inote{somewhere we need to link between passages and documents}
 
\subsection{Retrieval architectures}\label{ssec:architectures}
 
\begin{figure}[htb!]
    \ifigure{.90}{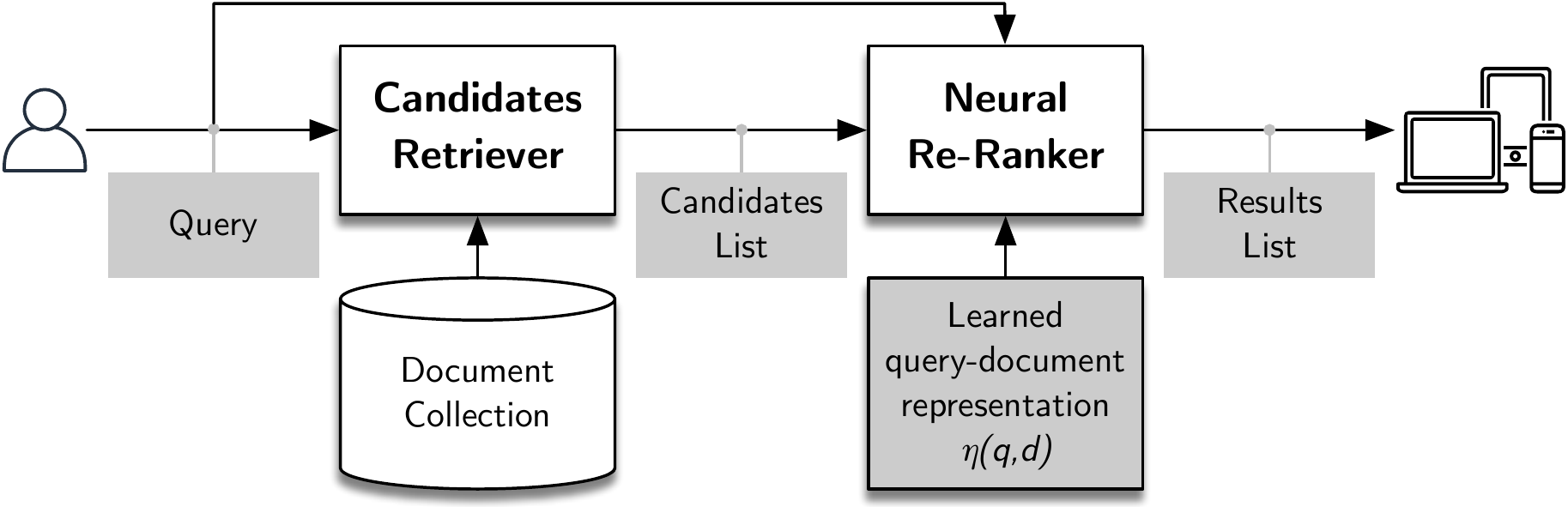}
    \caption{Re-ranking pipeline architecture for interaction-focused neural IR systems.}\label{fig:pipeline}
\end{figure}

Pre-trained language models successfully improve the  effectiveness of IR systems in the ad-hoc ranking task, but they are computationally very expensive. Due to these computational costs, the interaction-focused systems are not applied directly on the document collection, i.e., to rank all documents matching a query. They are deployed in a pipelined architecture (Figure~\ref{fig:pipeline}) by conducting first a preliminary ranking stage to retrieve a limited number of candidates, typically $1000$ documents, before re-ranking them with a more expensive neural re-ranking system, such as cross-encoders described in Section~\ref{sec:interaction}.

\begin{figure}[htb!]
    \ifigure{.80}{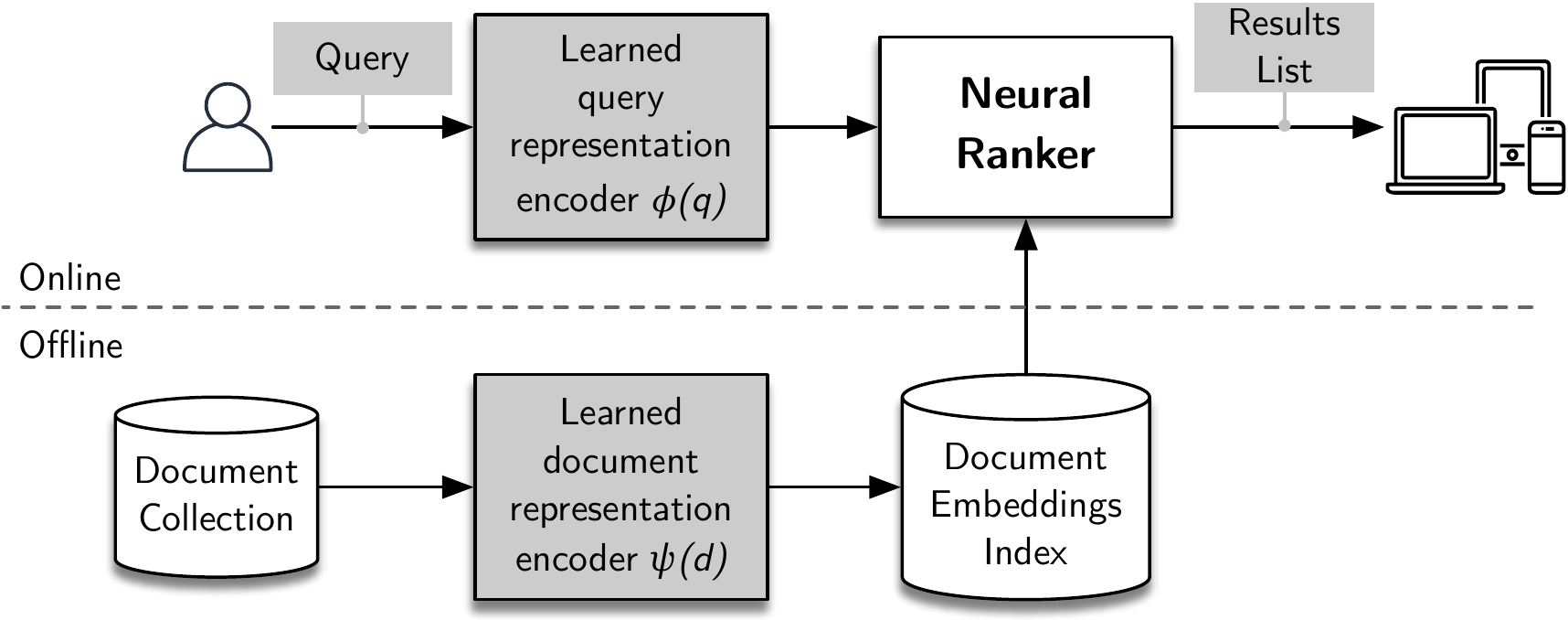}
    \caption{Dense retrieval architecture for representation-focused neural IR systems.}\label{fig:dense}
\end{figure}

The most important benefit of bi-encoders discussed in Section~\ref{sec:representation} is the possibility to pre-compute and cache the representations of a large corpus of documents with the learned document representation encoder $\psi(d)$. At query processing time, the learned query representation encoder must compute only the query representation $\phi(q)$, then the documents are ranked according to the inner product of their representation with the query embedding, and the top $k$ documents whose embeddings have the largest inner product w.r.t. the query embedding are returned to the user (Figure~\ref{fig:dense}).

\subsection{MIP and NN Search Problems}\label{ssec:search}
The pre-computed document embeddings are stored in a special data structure called \textit{index}. In its simplest form, this index must store the document embeddings and provide a search algorithm that, given a query embedding, efficiently finds the document embedding with the largest dot product, or, more in general, with the maximum inner product.

Formally, let $\phi \in \mathbb{R}^\ell$ denote a query embedding, and let $\Psi = \{\psi_1, \ldots, \psi_n\}$ denote a set of $n$ document embeddings, with $\psi_i \in \mathbb{R}^\ell$ for $i = 1, \ldots, n$. The goal of the \textit{maximum inner product (MIP) search} is to find the document embedding $\psi^* \in X$ such that
\begin{equation}\label{eq:mips}
    \psi^* = \arg \max_{\psi \in \Psi} \langle \phi, \psi \rangle
\end{equation}
A data structure designed to store $\Psi$ is called \textit{embedding index}. The na\"ive embedding index is the \textit{flat index}, which stores the document embeddings in $\Psi$ explicitly and performs an exhaustive search to identify $\psi^*$. Its complexity is $O(n\ell)$ both in space and time, so it is particularly inefficient for large $n$ or $\ell$ values.

A common approach to improve the space and time efficiency of the flat index is to convert the maximum inner product search into a \textit{nearest neighbour (NN) search}, whose goal is to find the document embedding $\psi^\dagger$ such that
\begin{equation}\label{eq:nns}
    \psi^\dagger = \arg \min_{\psi \in \Psi} \|\phi - \psi\|
\end{equation}

Many efficient index data structures exist for NN search. To leverage them with embedding indexes, MIP search between embeddings must be adapted to use the Euclidean distance and NN search. This is possible by applying the following transformation from $\mathbb{R}^\ell$ to $\mathbb{R}^{\ell+1}$~\cite{10.5555/3045118.3045323,xbox}:
\begin{equation}
\label{eq:transformation}
    \hat{\phi} = \begin{bmatrix}\phi/\|\phi\| \\ 0\end{bmatrix},\quad
    \hat{\psi} = \begin{bmatrix}\psi/M \\ \sqrt{1 - \|\psi\|^2/M^2}\end{bmatrix},
\end{equation}
where $M = \max_{\psi \in \Psi} \|\psi\|$. By using this transformation,
%the maximization problem of the inner product $\langle q,x \rangle$ becomes exactly equivalent to the minimization problem of the distance $\|\hat{q} - \hat{x}\|$. In fact,
the MIP search solution $\psi^*$ coincides with the NN search solution $\hat{\psi}^\dagger$. In fact,
%assuming without loss of generality that $\|\psi_a\|=1$,
we have:
\begin{equation*}
    \min \|\hat{\phi} - \hat{\psi}\|^2 =
    \min \big( \|\hat{\phi}\|^2 + \|\hat{\psi}\|^2 - 2 \langle \hat{\phi}, \hat{\psi} \rangle \big) =
    \min \big(2 - 2\langle \phi, \psi/M \rangle \big) =
    \max \langle \phi, \psi\rangle.
\end{equation*}
Hence, hereinafter we consider the MIP search for ranking with a dense retriever as a NN search based on the Euclidean distance among the transformed embeddings $\hat{\phi}$ and $\hat{\psi}$ in $\mathbb{R}^{\ell+1}$. 
%Intuitively, assuming $l = 2$, the transformation~\eqref{eq:transformation} maps arbitrary \textit{query and document} vectors in $\mathbb{R}^2$ into unit-norm \textit{query and document} vectors in $\mathbb{R}^3$, i.e., the transformed vectors are mapped on the surface of the unit sphere in $\mathbb{R}^3$.
To simplify the notation, from now on we drop the hat symbol from the embeddings, i.e.,  $\hat{\phi} \to \phi$ and $\hat{\psi} \to \psi$, and we consider $\ell+1$ as the new dimension $\ell$, i.e., $\ell+1 \to \ell$.

The index data structures for exact NN search in low dimensional spaces have been very successful, but they are not efficient with high dimensional data, as in our case, due to the curse of dimensionality. It is natural to make a compromise between search accuracy and search speed, and the most recent search methods have shifted to \textit{approximate nearest neighbor (ANN) search}.

The ANN search approaches commonly used in dense retrieval can be categorised into three families: locality sensitive hashing approaches, quantisation approaches, and graph approaches.

\subsection{Locality sensitive hashing approaches}\label{ssec:lsh}

\textit{Locality sensitive hashing (LSH)}~\cite{10.1145/276698.276876} is based on the simple idea that, if two embeddings are close together, then after a ``projection'', using an hash function, these two embeddings will remain close together. 
This requires that:
\begin{itemize}
\item for any two embeddings $\psi_1$ and $\psi_2$ that are close to each other, there is a high probability $p_1$ that they fall into the same hash bucket;
\item for any two embeddings $\psi_1$ and $\psi_2$ that are far apart, there is a low probability $p_2 < p_1$ that they fall into the same hash bucket.
\end{itemize}
The actual problem to solve is to design a family of LSH functions fulfilling these requirements. LSH functions have been designed for many distance metrics. For the euclidean distance, a popular LSH function $h(\psi)$ is the \textit{random projection}~\cite{10.1145/997817.997857}.%, defined as follows:
%\begin{equation}\label{eq:lsh}
%    h_{a,b}(\psi) = \bigg\lfloor \frac{\langle a, \psi\rangle + b}{c} \bigg\rfloor
%\end{equation}
%where $a \in \mathbb{R}^\ell$ is a vector whose components are drawn independently from the normal distribution, and $b \in \mathbb{R}$ is uniformly drawn from $[0, c]$ ($c$ is a quantisation parameter depending on the data). 
A set of random projections defines a family of hash functions $\mathcal{H}$ that can be used to build a data structure for ANN search.
First, we sample $m$ hash functions $h_1(\psi), \ldots, h_m(\psi)$ independently and uniformly at random from $\mathcal{H}$, and we define the function family $\mathcal{G} = \{g : \mathbb{R}^\ell \to \mathbb{Z}^m\}$, where $g(\psi) = \big(h_1(\psi),\ldots,h_m(\psi)\big)$, i.e., $g$ is the concatenation of $m$ hash functions from $\mathcal{H}$. Then, we sample $r$ functions $g_1(\psi), \ldots, g_r(\psi)$ independently and uniformly at random from $\mathcal{G}$, and each function $g_i$ is used to build a hash table $H_i$. %By using $m$ hash functions from $\mathcal{H}$, we reduce the low probability $p_2$ of a single hash function to $p_2^m$, but we also reduce the high probability $p_1$ to $p_1^m$. The $l$ hash tables are needed to increase the high probability 

Given the set of document embeddings $\Psi$ and selected the values of the parameters $r$ and $m$, % and $c$, 
an \textit{LSH index} is composed of $r$ hash tables, each containing $m$ concatenated random projections. For each $\psi \in \Psi$, $\psi$ is inserted in the $g_i(\psi)$ bucket %\footnote{Since the total number of hash buckets may be large, only non-empty buckets are retained using regular hashing.}
for each hash table $H_i$, for $i=1, \ldots, r$.
At query processing time, given a query embedding, we first generate a candidate set of document embeddings by taking the union of the contents of all $r$ buckets in the $r$ hash tables the query is hashed to. The final NN document embedding is computed performing an exhaustive exact search within the candidate set.

The main drawback of the LSH index is that it may require a large number of hash tables to cover most nearest neighbors, and it requires to store the original embeddings to perform the exhaustive exact search. Although some optimisations have been proposed~\cite{falconn}, the space consumption may be prohibitive with very large data sets.
%Using LSH, the set of document embeddings $X$ is pre-processed to compute, for each $x \in X$, the $l$ $k$-dimensional buckets $g_1(x), \ldots, g_l(x)$. To process a query embedding $q$, the query is first hashed into $l$ buckets using the same hash functions used for document embeddings. Then, the document embeddings lying in these buckets represent the embedding ``closest'' to the query in the projected space. 

\subsection{Vector quantisation approaches}\label{ssec:pq}

Instead of random partitioning the input space $\Psi$ as in LSH, the input space can be partitioned according to the data distribution. By using the $k$-means clustering algorithm on $\Psi$ we can compute $k$ \textit{centroids} $\mu_1, \ldots, \mu_k$, with $\mu_i \in \mathbb{R}^\ell$ for $i = 1, \ldots, k$ that can be used to partition the input space $\Psi$. The set $M = \{\mu_1, \ldots, \mu_k\}$ is called a \textit{codebook}. Given a codebook $M$, a \textit{vector quantiser} $q: \mathbb{R}^\ell \to \mathbb{R}^\ell$ maps a vector $\psi$ to its closest centroid:
%Given $k$ centroid embeddings $M = \{\mu_1, \ldots, \mu_k\}$ over $X \subset \mathbb{R}^d$, we can define a \textit{quantiser} as a function $q_M: \mathbb{R}^d \to \mathbb{R}^d$ such that:
\begin{equation}\label{eq:quantiser}
    q(\psi) = \arg \min_{\mu \in M} \|\psi - \mu\|
\end{equation}
Given a codebook $M$, an \textit{IVF} (Inverted File) \textit{index} built over $M$ and $\Psi$ stores the set of document embeddings $\Psi$ in $k$ \textit{partitions} or \textit{inverted lists}  $L_1, \ldots, L_k$, where $L_i = \{\psi \in \Psi : q(\psi) = \mu_i\}$. At query processing time, we specify to search for the NN document embeddings in $p > 0$ partitions. If $p=k$, the search is exhaustive, but if $p < k$, the search is carried out in the partitions whose centroid is closer to the query embedding. In doing so the search is not guaranteed to be exact, but the search time can be sensibly reduced. In fact, an IVF index does not improve the space consumption, since it still needs to store all document embeddings, but it can reduce the search time depending on the number of partitions processed for each query.

A major limitation of IVF indexes is that they can require a large number of centroids~\cite{Gersho:1992:VQS}. To address this limitation, \textit{product quantisation}~\cite{pq} divides each vector $\psi \in \Psi$ into $m$ \textit{sub-vectors} $\psi = [\psi_1 | \psi_2 | \cdots | \psi_m]$. Each sub-vector $\psi_j \in \mathbb{R}^{\ell/m}$ with $j = 1,\ldots,m$ is quantised independently using its own \textit{sub-vector quantiser} $q_j$. Each vector quantiser $q_j$ has its own codebook $M_j = \{\mu_{j,1}, \ldots, \mu_{j,k}\}$. Given the codebooks $M_1, \ldots, M_m$, a \textit{product quantiser} $pq: \mathbb{R}^\ell \to \mathbb{R}^\ell$ maps a vector $\psi$ into the concatenation of the centroids of its sub-vector quantisers:
\begin{equation}\label{eq:pquantiser}
    pq(\psi) = [ q_1(\psi_1) | q_2(\psi_2) | \cdots | q_m(\psi_m)] = [\mu_{1, i_1}|\mu_{2, i_2}|\ldots|\mu_{m, i_m}]
\end{equation}
Note that a product quantiser can output any of the $k^m$ centroid combinations in $M_1 \times \ldots \times M_m$. 

A \textit{PQ} (Product Quantization) \textit{index} stores, for each embedding $\psi \in \Psi$, its encoding $i_1,\ldots,i_m$, that requires $m\log k$ bits of storage. 
At query processing time, the document embeddings are processed exhaustively. However the distance computation between a query embedding $\phi$ and a document embedding $\psi$ is carried out using the product quantisation of the document embedding $pq(x)$:
\begin{equation}\label{eq:adc}
    \|\psi-\phi\|^2 \approx \|pq(\psi) - \phi\|^2 = \sum_{j=1}^m \|q_j(\psi_j) - \phi_j\|^2
\end{equation}
To implement this computation, $m$ lookup tables are computed, one per sub-vector quantiser: the $j$-th table is composed of the squared distances between the $j$-th sub-vector of $\phi$, and the centroids of $M_j$. These tables can be used to quickly compute the sums in Eq.~\eqref{eq:adc} for each document embedding.

ANN search on a PQ index is fast, requiring only $m$ additions, and memory efficient, but it is still exhaustive. To avoid it, an \textit{IVFPQ index} exploits inverted files and product quantisation jointly. Firstly, a \textit{coarse quantiser} partitions the input dataset into inverted lists, for a rapid access to small portions of the input data. In a given inverted list, the difference between each input data and the list centroid, i.e., the input \textit{residual}, is encoded with a product quantiser. In doing so, the exhaustive ANN search can be carried out only in a limited number of the partitions computed by the coarse quantiser.

\subsection{Graph approaches}\label{ssec:hnsw}

The distances between vectors in a dataset can be efficiently stored in a graph-based data structure called \textit{$k$NN graph}. In a $k$NN graph $G=(V,E)$, each input data $\psi \in \Psi$ is represented as a node $v \in V$, and, for its $k$ nearest neighbours, a corresponding edge is added in $E$. The computation in an exact $k$NN graph requires $O(n^2)$ similarity computation, but many approximate variants are available~\cite{10.1145/1963405.1963487}. To search for an approximate nearest neighbour to an element $\phi$ using a $k$NN graph, a \textit{greedy heuristic search} is used. Starting from a predefined entry node, the graph is visited one node at a time, keeping on finding the closest node to $\phi$ among the unvisited neighbour nodes. The search terminates when there is no improvement in the current NN candidate. In practice, several entry nodes are used together with a search budget to avoid local optima. For a large number of nodes, the greedy heuristic search on the $k$NN graph becomes inefficient, due to the long paths potentially required to connect two nodes. Instead of storing only short-range edges, i.e., edges connecting two close nodes, the $k$NN graph can be enriched with randomly generated long-range edges, i.e., edges connecting two randomly-selected nodes. This kind of $k$NN graph is a \textit{navigable small world (NSW) graph}~\cite{nsw}, for which the greedy search heuristic is theoretically and empirically efficient~\cite{smallworld}.

A \textit{hierarchical NSW (HNSW) index} stores the input data into multiple NSW graphs. The bottom layer graph contains a node for each input element, while the number of nodes in the other graphs decreases exponentially at each layer. The search procedure for approximate NN vectors starts with the top layer graph. At each layer, the greedy heuristic searches for the closest node, then the next layer is searched, starting from the node corresponding to the closest node identified in the preceding graph. At the bottom layer, the greedy heuristic searches for the $k$ closest nodes to be returned~\cite{hnsw}.

\subsection{Optimisations}\label{ssec:optimisations}

Implementations of the embedding indexes presented in the previou sections are available in many open-source production-ready search engines such as Lucene\footnote{\url{https://lucene.apache.org}} and Vespa\footnote{\url{https://vespa.ai}}. In the IR research community, FAISS is the most widely adopted framework for embedding indexes~\cite{faiss}. Among others, FAISS includes implementations of flat, LSH, IVF, PQ, IVFPQ and HNSW indexes. 

Single representation systems such as \dpr~\cite{dpr}, \ance~\cite{ance}, and \star~\cite{star} use flat indexes. In these cases, it is unfeasible to adopt product quantisation indexes due to their negative impact on IR-specific metrics, mainly caused by the separation between the document encoding and embedding compression phases. To overcome this limitation, several recent techniques such as {\sf Poemm}~\cite{poeem}, {\sf JPQ}~\cite{jpq} and {\sf RepCONC}~\cite{repconc} aim to train at the same time both phases. In doing so, during training, these techniques learn together the query and document encoders together while performing product quantisation.

\begin{figure}[htb!]
    \ifigure{.99}{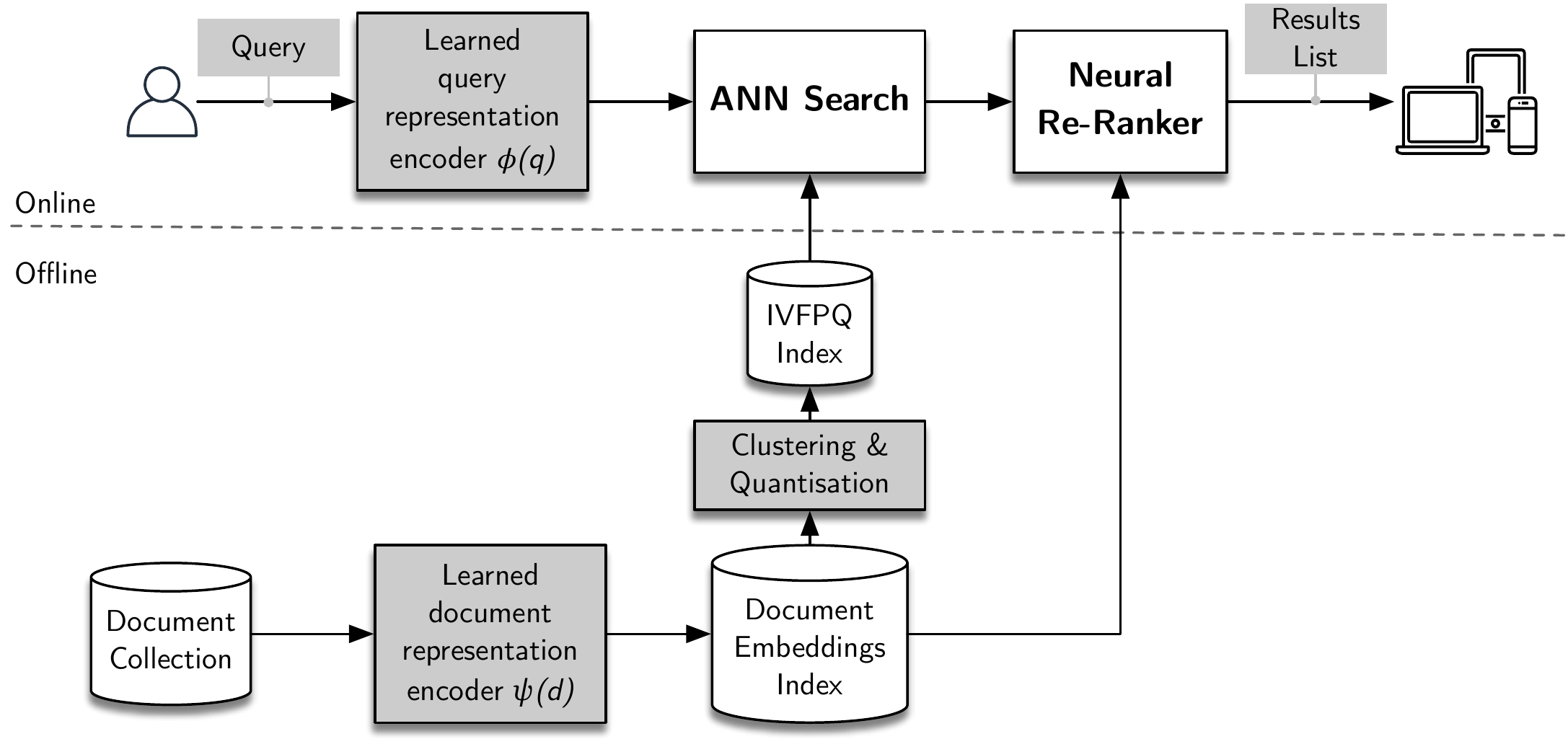}
    \caption{Ranking pipeline architecture for multiple representation systems.}\label{fig:colbert}
\end{figure}

Multiple representation systems such as \colbert~\cite{colbert} are characterised by a very large number of document embeddings. They do not use flat indexes, due to unacceptable efficiency degradation of brute-force search, and exploit IVFPQ indexes and ANN search.
With these indexes, the document embeddings are stored in a quantised form, suitable for fast searching. However, the approximate similarity scores between these compressed embeddings are inaccurate, and hence are not used for computing the final top documents. Indeed, in a first stage, ANN search computes, for each query embedding, the set of the $k'$ most similar document embeddings; the retrieved document embeddings for each query embedding are mapped back to their documents. These documents are exploited to compute the final list of top $k$ documents in a second stage. To this end, the set of documents computed in the first stage is re-ranked using the query embeddings and the documents' multiple embeddings to produce exact scores that determine the final ranking, according to the relevance function in Eq.~\eqref{eq:maxsim2} (see Figure~\ref{fig:colbert}. 

Further optimisations can reduce the number of query embeddings to be processed in the first stage~\cite{eaat}, or the number of documents to be processed in the second stage~\cite{approx}.

\section{Learned Sparse Retrieval}\label{sec:sparse}

Traditional IR systems are based on sparse representations, inverted indexes and lexical-based relevance scoring functions such as BM25. In industry-scale web search, BM25 is a widely adopted baseline due to its trade-off between effectiveness and efficiency. On the other side, neural IR systems are based on dense representations of queries and documents, that have shown impressive benefits in search effectiveness, but at the cost of query processing times. In recent years, there have been some proposals to incorporate the effectiveness improvements of neural networks into inverted indexes, with their efficient query processing algorithms, through \textit{learned sparse retrieval} approaches.
%Instead of computing dense representation of documents, 
In learned sparse retrieval the transformer architectures are used in different scenarios:
\begin{itemize}
    \item \textit{document expansion learning}: sequence-to-sequence models are used to modify the actual content of documents, boosting the statistics of the important terms and generating new terms to be included in a document;
    \item \textit{impact score learning}: the output embeddings of documents provided as input to encoder-only models are further transformed with neural networks to generate a single real value, used to estimate the average relevance contribution of the term in the document;
    \item \textit{sparse representation learning}: the output embeddings of documents provided as input to encoder-only models are projected with a learned matrix on the collection vocabulary, in order to estimate the relevant terms in a document. These relevant terms can be part of the documents or not, hence representing another form of document enrichment. 
\end{itemize}

In Sections~\ref{ssec:docexp}, \ref{ssec:impact}, and \ref{ssec:splade},
%In the following, we will 
we describe the main existing approaches in these scenarios, respectively.

\subsection{Document expansion learning}\label{ssec:docexp}

Document expansion techniques address the vocabulary mismatch problem~\cite{zhao2012modeling}: queries can use terms semantically similar but lexically different from those used in the relevant documents. Traditionally, this problem has been addressed using query expansion techniques, such as relevance feedback~\cite{rocchio1971relevance} and pseudo relevance feedback~\cite{lavrenko2001relevance}. The advances in neural networks and natural language processing have paved the way to different techniques to address the vocabulary mismatch problem by expanding the documents by learning new~terms.

\begin{figure}[htb!]
    \ifigure{1}{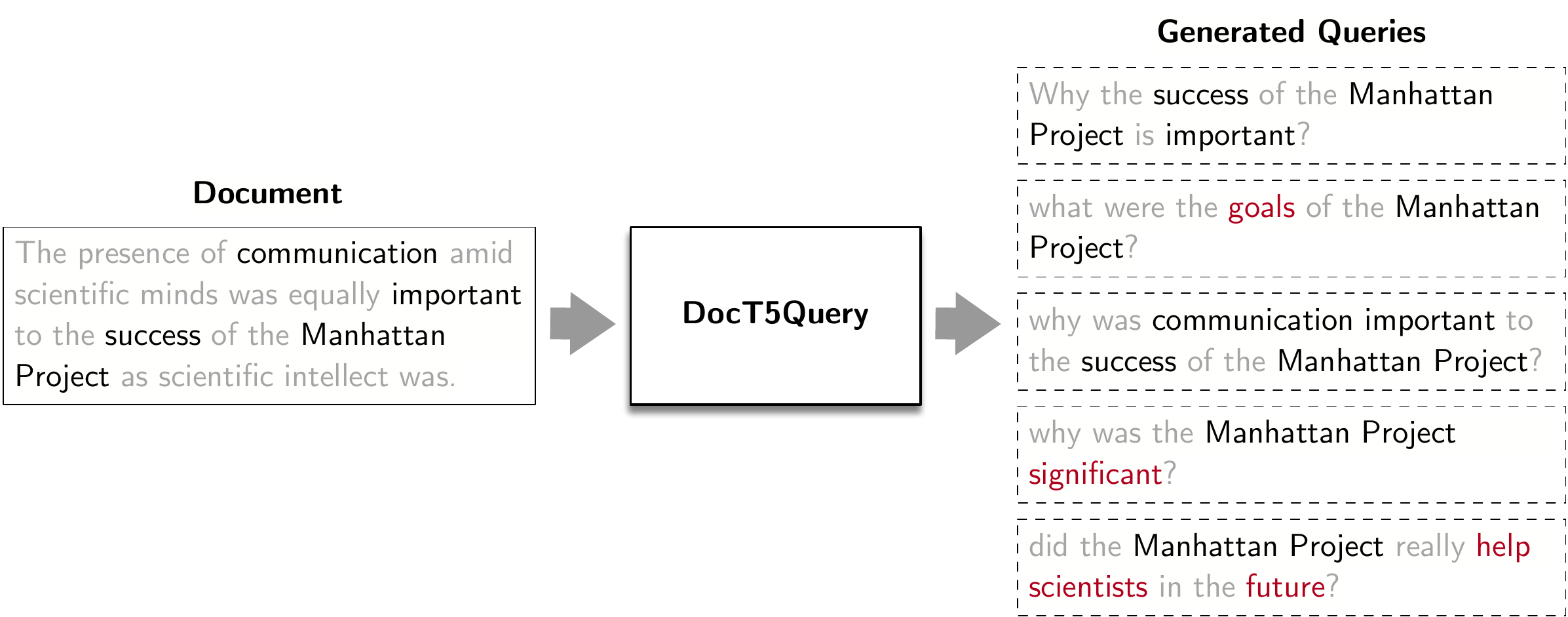}
    \caption{Example of generated queries using \dtfiveq. Black terms denote boosted important terms, red terms denote new important terms not present in the original document. }\label{fig:dt5q}
\end{figure}

\dtoq~\cite{doc2query} and \dtfiveq~\cite{docTTTTTquery} showed for the first time that transformer architectures can be used to expand the documents' content to include new terms or to boost the statistics of existing termw. Both approaches focus on the same task, that is, generating new queries for which a specific document will be relevant. Given a dataset of query and relevant document pairs, \dtoq\ fine-tunes a sequence-to-sequence transformer model~\cite{transformer}, while \dtfiveq\ fine-tunes the \tfive\ model~\cite{t5} by taking as input the relevant document and generating the corresponding query. Then, the fine-tuned model is used to predict new queries using top $k$ random sampling~\cite{fan-etal-2018-hierarchical} to enrich the document by appending these queries before indexing, as illustrated in Figure~\ref{fig:dt5q}.

Instead of leveraging the encoder-decoder models for sentence generation and fine-tune them on document expansion, a different approach computes the importance of all terms in the vocabulary w.r.t. a given document and selects the most important new terms to enrich the document, leveraging an encoder-only architecture to compute the document embeddings.
\tilde~\cite{tildetwo} exploits the \bert\ model to compute the \cls\ output embedding of a document, and linearly projects it over the whole \bert\ vocabulary. In doing so, \tilde\ computes a probability distribution over the vocabulary, i.e., a document language model, and then adds to the document a certain number of new terms, corresponding to those with the highest probabilities. As another way of expanding documents, \sparterm~\cite{bai2020sparterm} computes a document language model for each \bert\ output token, including \cls, and sums them up to compute the term importance distribution over the vocabulary for the given document. Finally, a learned \textit{gating mechanism} only keeps a sparse subset of those, to compute the final expanded document contents.

\subsection{Impact score learning}\label{ssec:impact}

Classical inverted indexes store statistical information on term occurrences in documents in posting lists, one per term in the collection. Every posting list stores a posting for each document in which the corresponding term appears in, and the posting contains a document identifier and the in-document term frequency, i.e., a positive integer counting the number of occurrences of the term in the document. When a new query arrives, the posting lists of the terms in the query are processed to compute the top scoring documents, using a classical ranking function, such as BM25, and efficient query processing algorithms~\cite{fntir}. 

The goal of impact score learning is to leverage the document embeddings generated by an encoder-only model to compute a single integer value to be stored in postings, and to be used as a proxy of the relevance of the term in the corresponding posting, i.e., its \textit{term importance}.
The simplest way to compute term importance in a document is to project the document embeddings of each term with a neural network into a single-value representation, filtering out negative values with \relu\ functions and discarding zeros. To save space, the real values can be further quantised into a 8-bit positive integers.
%Each posting list stores information about a term properties in a collection, i.e., for each term in the collection vocabulary. 
A common problem in impact score learning is the vocabulary to use. Since most encoder-only models use a sub-word tokeniser, the collection vocabulary can be constructed in two different ways: by using the terms produced by the encoder-specific sub-word tokeniser, e.g., by \bert-like tokenisers, or by using the terms produced by a word tokeniser. These two alternatives have an impact on the final inverted index: in the former case, we have fewer terms, but longer and denser posting lists, while in the latter case, we have more terms, with shorter posting lists and with smaller query processing times~\cite{deepimpactv2}.

The current impact score learning systems are \deepct~\cite{deepct,deepct2}, \deepimpact~\cite{deepimpactv1}, \tilde~\cite{tilde,tildetwo}, and \unicoil~\cite{unicoil}.

\deepct~\cite{deepct,deepct2} represents the first example of term importance boosting. \deepct\ exploits the contextualised word representations from \bert\ to learn new in-document term frequencies, to be used with classical ranking functions such as BM25. For each term $w_i \in \mathcal{V}$ in a given document, \deepct\ estimates its context-specific \textit{importance} $z_i \in \mathbb{R}$, that is then scaled and rounded as frequency-like integer value $tf_i$ that can be stored in an inverted index.
Formally, for each document $d \in \mathcal{D}$, \deepct\ projects the $\ell$-dimensional representations $\psi_i$ for each input \bert\ token $w_i$ in the document, with $i=1,\ldots,|d|$, into a scalar term importance with the learned matrix $W \in \mathbb{R}^{1 \times \ell}$:
\begin{equation}
    \begin{split}
        [\psi_0, \psi_1, \ldots] &= {\encoder}(d)\\
        %[z_0, z_1, \ldots] &= [W\psi_0, W\psi_1, \ldots]\\
        z_i &= W\psi_i\\
    \end{split}
\end{equation}
\deepct\ is trained with a per-token regression task, trying to predict the importance of the terms. The actual term importance to predict is derived from the document containing the term, or from a training set of query, relevant document pairs. A term appearing in multiple relevant documents and in different queries has a higher importance than a term matching fewer documents, and/or fewer queries. 
To handle \bert's sub-word tokens, \deepct\ uses the importance of the first sub-word token for the entire word, and when a term occurs multiple times in the document, it takes the maximum importance across the multiple occurrences.

\deepimpact~\cite{deepimpactv1} proposes for the first time to directly compute an impact score for each unique term in a document, without resorting to classical ranking functions, but simply summing up, at query processing time, the impacts of the query terms appearing in a document to compute its relevance score.
For each term $w_i \in \mathcal{V}$ in a given document $d \in \mathcal{D}$, \deepimpact\ estimates its context-specific \textit{impact} $z_i \in \mathbb{R}$. \deepimpact\ feeds the encoder-only model with the document sub-word tokens, producing an embedding for each input token. A non-learned \textit{gating layer} {\sf Mask} removes the embeddings of the sub-word tokens that do not correspond to the first sub-token of the whole word. Then \deepimpact\ transforms the remaining $\ell$-dimensional representations with two feed forward networks with \relu\ activations. The first network has a weight matrix $W_1 \in \mathbb{R}^{\ell \times \ell}$, and the second network has a weight matrix $W_2 \in \mathbb{R}^{1 \times \ell}$:
\begin{equation}
    \begin{split}
        [\psi_0, \psi_1, \ldots] &= {\encoder}({\dtfiveq}(d))\\
        [x_0, x_1, \ldots] &= {\sf Mask}(\psi_0, \psi_1, \ldots)\\
        %[z_0, z_1, \ldots] &= [{\sf ReLU}(W_1\psi_0), {\sf ReLU}(W_1\psi_1), \ldots]\\
        y_i &= {\relu}(W_1 x_i)\\
        z_i &= {\relu}(W_2 y_i)\\
    \end{split}
\end{equation}
The output real numbers $z_i$, with $i=1,\ldots,|d|$, one per whole word in the input document, are then linearly quantised into 8-bit integers that can be stored in an inverted index. This produces a single-value score for each unique term in the document, representing its impact. Given a query $q$, the score of the document $d$ is simply the sum of impacts for the intersection of terms in $q$~and~$d$.
\deepimpact\ is trained with query, relevant document, non-relevant document triples,
%The model converts each document into a list of scores, corresponding to the document terms matching the query. These scores are then summed up, obtaining an accumulated query-document score. 
and, for each triple, two scores for the corresponding two documents are computed. The model is optimized via pairwise cross-entropy loss over the document scores.
Moreover, \deepimpact\ has been the first sparse learned model leveraging at the same time documents expansion learning and impact score learning. In fact, \deepimpact\ leverages \dtfiveq\ to enrich the document collection before learning the term impact.

\tilde~\cite{tildetwo} computes the terms' impact with an approach similar to \deepimpact. The main differences are (i) the use of a single layer feed forward network with \relu\ activations, instead of a two-layer network, to project the document embeddings into a single positive scalar value using a learned matrix $W \in \mathbb{R}^{1 \times \ell}$, (ii) the use of its own document expansion technique, as discussed in Section~\ref{ssec:docexp}, (iii) the use of an index with sub-word terms instead of whole word terms, and (iv) the selection of the highest-valued impact score for a token if that token appears multiple times in a document:
\begin{equation}
    \begin{split}
        [\psi_0, \psi_1, \ldots] &= {\encoder}({\tilde}(d))\\
        z_i &= {\relu}(W \psi_i)\\
    \end{split}
\end{equation}
The $z_i$ scores are then summed up, obtaining an accumulated query-document score. 

\unicoil~\cite{unicoil} exploits the \coil\ approach (see Sec.~\ref{sec:representation}), but instead of projecting the query and document embeddings on 8-32 dimensions, it projects them to single-dimension query weights and document weights. 
In \unicoil\ the query and document \cls\ embeddings are not used, and the embeddings corresponding to normal query and document tokens are projected into single scalar values $v_1, \ldots, v_{|d|}$ using a learned matrix $W \in \mathbb{R}^{1 \times \ell}$, with \relu\ activations on the output term weights of the base \coil\ model to force the model to generate non-negative weights.

\begin{equation}\label{eq:maxsim4}
    \begin{split}
        [\phi_0, \phi_1, \ldots] &= {\encoder}(q)\\
        [\psi_0, \psi_1, \ldots] &= {\encoder}({\dtfiveq}(d))\\
        [v_1, v_2, \ldots] &= [W\phi_1, W\phi_2, \ldots]\\
        [z_1, z_2, \ldots] &= [W\psi_1, W\psi_2, \ldots]\\
        s(q,d) &= \sum_{t_i \in q}\max_{t_j \in d, \, t_j = t_i} v_i  z_j
    \end{split}
\end{equation}

The document weights $z_i$ are then linearly quantised into 8-bit integers, and the final query-document score is computed by summing up the highest valued document impact scores times its query weight $v_i$, computed at query processing time, as in Eq.~\eqref{eq:maxsim4}.

\subsection{Sparse representation learning}\label{ssec:splade}

Instead of independently learning to expand the documents and then learning the impact score of the terms in the expanded documents, sparse representation learning aims at learning both at the same time.
At its core, sparse representation learning projects the output embeddings of an encoder-only model into the input vocabulary, compute, for each input term in the document, a language model, i.e., a probability distribution over the whole vocabulary. These term-based language models capture the semantic correlations between the input term and all other terms in the collection, and they can be used to (i) \textit{expand} the input text with highly correlated terms, and (ii) \textit{compress} the input text by removing terms with low probabilities w.r.t. the other terms.

Encoder-only models such as \bert\ already compute term-based language models, as part of their training as masked language models. Formally, given a document $d$, together with the output embeddings $\psi_{\cls}, \psi_1, \ldots, \psi_{|d|}$, an encoder-only model also returns the \textit{masked language heads} $\chi_1, \ldots, \chi_{|d|}$, one for each token in the document, where $\chi_i \in \mathbb{R}^{|\mathcal{V}|}$ for $ i = 1, \ldots, |d|$ is an estimation of the importance of each word in the vocabulary implied by the $i$-th token in the document $d$.
\epic~\cite{epic} and \sparterm~\cite{bai2020sparterm} have been the first systems focusing on vocabulary-based expansion and importance estimation, and inspired the \splade~\cite{splade} system, on which we focus.

For a given document $d \in \mathcal{D}$, \splade\ computes its per-token masked language heads $\chi_1, \ldots, \chi_{|d|}$ using \bert, filters and sums up these vocabulary-sized vectors into a single vector $\gamma(d) \in \mathbb{R}^{|\mathcal{V}|}$ representing the whole document, and then uses this vector to represent the document itself, together with the term importance scores:
\begin{equation}\label{eq:splade1}
    \begin{split}
        [\chi_1, \ldots, \chi_{|d|}] &= {\encoder}(d)\\
        \gamma(d) &= \sum_{i=1}^{|d|} \log \big(1+ {\relu}(\chi_i)\big)\\
    \end{split}
\end{equation}
The logarithm and \relu\ functions in Eq.~\eqref{eq:splade1} are computed element-wise; the logarithm prevents some terms with large values from dominating, and the \relu\ function deals with the negative components of $\gamma(d)$.

The document representation $\gamma$ potentially contains all terms in the vocabulary, even if the logarithm and \relu\ functions in Eq.~\eqref{eq:splade1} can zero out some of its components. To learn to ``sparsify'' the document representations, \citet{splade} leverage the \flops\ regulariser $\mathcal{L}_{\sf FLOPS}$~\cite{flops}.
As part of the \splade\ loss function used during training, the FLOPS loss is computed as the sum, across the terms in the vocabulary, of the squared probability $p_w^2$ that a term $w$ has a non-zero weight in a document. Minimising the \flops\ loss coincides with minimising the non-zero weights in a document, i.e., maximising the number of zero weights in a document. The square operation helps in reducing high term weights more than low term weights. The probability that a term $w \in \mathcal{V}$ has a non-zero weight in a document $d$ is proportional to the average weight of that term $\gamma_t(d)$ estimated through the whole collection. To make the computation feasible, the average is computed on a batch $b$ of documents during training, considered as a representative sample of the whole collection:
\begin{equation}\label{eq:flops}
    \mathcal{L}_{\flops} = \sum_{t \in V} p_t^2 = \sum_{t \in V} \bigg(\frac{1}{|b|}\sum_{d \in b} \gamma_t(d)\bigg)^2
\end{equation}

\splade\ does not limit expansion to documents only. Indeed, Eq.~\eqref{eq:splade1} can be applied to a query $q$ as well, to compute the corresponding vector $\gamma(q) \in \mathbb{R}^{|\mathcal{V}|}$. However, this query expansion must be carried out at query processing time; to reduce the latency, the expanded query should be far more sparse than a document. To enforce this different behaviour, \citet{splade} adopt two distinct \flops\ regularisers for documents and queries, both as in Eq.~\ref{eq:flops}.

%A main limitation of DeepCT lies in the fact that it does not address the vocabulary mismatch problem~\cite{zhao2012modeling}: only terms already appearing in the documents will receive learned weights to improve their relevance signals. A way to address this limitations was first proposed by \deepimpact. Instead of using the original document collection, the documents can be expanded to include new terms to address the vocabulary mismatch. Leveraging sequence-to-sequence models, document expansion techniques such as \doctfquery~\cite{docTTTTTquery} and \tildetwo~\cite{tilde} enrich the documents with new terms, as well as modifying the occurrences of existing terms. In details, \doctfquery\ exploits the T5 casual language model to predict queries potentially relevant to a given document, and enriches the document by appending these queries before indexing. \tildetwo\ exploits the BERT contextualized language model to compute document likelihood models used to enrich the documents with new terms.  As another way of expanding documents, \sparterm~\cite{bai2020sparterm} predicts an importance score for every term in the vocabulary and uses a gating mechanism to only keep a sparse subset of those. However, experiments show that the benefits of \sparterm\ are very similar to those of \doctfquery

%\subsection{Knowledge Distillation}

%\subsection{Hybrid Approaches}

%\subsection{Efficiency Considerations}

%\section{Other Applications}

\section{Conclusions}\label{sec:conc}

This overview aimed to provide the foundations of neural IR approaches for ad-hoc ranking, focusing on the core concepts and the most commonly adopted neural architecture in current research. In particular, in Section~\ref{sec:math} we provided a background on the different ways to represent texts, such as queries and documents, to be processed in IR systems. Sections~\ref{sec:interaction}~and~\ref{sec:representation} illustrated the main system architectures in neural IR, namely interaction-focused and representation-focused systems. In Section~\ref{sec:dense} we reviewed the multi-stage retrieval architectures exploiting neural IR systems, and we provided a quick overview of the embedding indexes and algorithms for dense retrieval. Finally, in Section~\ref{sec:sparse}, we discussed the main state-of-the-art solutions for learning sparse representations. 

Neural IR systems are currently a hot research topic, and many excellent surveys complement and deepen the concepts discussed in this overview. The interested readers can find a fully-fledged detailed presentation of pre-trained transformers for ranking, with many experimental evaluations, in the recent book by~\cite{lin2020pretrained}, as well as applications of dense retrieval to conversational system in the books by \citet{10.1145/3209978.3210183} and \citet{https://doi.org/10.48550/arxiv.2201.08808}.
\newpage

\bibliography{biblio}

\end{document}